\documentclass[12pt,a4paper,english,superscriptaddress,aps,preprint]{revtex4}
\usepackage{amsmath}
\usepackage{amssymb}
\usepackage{graphicx}
\usepackage{hhline}
\usepackage{textcomp}
\makeatletter
\usepackage{babel}
\newcommand{\bea}{\begin{eqnarray}}
\newcommand{\eea}{\end{eqnarray}}

\newcommand{\pa}{\partial}
\renewcommand{\a}{\alpha}

\renewcommand{\b}{\beta}

\newcommand{\q}{\theta}

\newcommand{\be}{\begin{equation}}
\newcommand{\ee}{\end{equation}}
\begin{document}
\immediate\write16{<<WARNING: LINEDRAW macros work with emTeX-dvivers
                    and other drivers supporting emTeX \special's
                    (dviscr, dvihplj, dvidot, dvips, dviwin, etc.) >>}

%% Macros for drawing Feynman graphs and other complex diagrams
%% Designed by A.V.Voronin (1993); modified in 1995
%% Steklov Math. Inst., e-mail: av@voronin.mian.su
%%
\newdimen\Lengthunit       \Lengthunit  = 1.5cm
\newcount\Nhalfperiods     \Nhalfperiods= 9
\newcount\magnitude        \magnitude = 1000

\catcode`\*=11
\newdimen\L*   \newdimen\d*   \newdimen\d**
\newdimen\dm*  \newdimen\dd*  \newdimen\dt*
\newdimen\a*   \newdimen\b*   \newdimen\c*
\newdimen\a**  \newdimen\b**
\newdimen\xL*  \newdimen\yL*
\newdimen\rx*  \newdimen\ry*
\newdimen\tmp* \newdimen\linwid*

\newcount\k*   \newcount\l*   \newcount\m*
\newcount\k**  \newcount\l**  \newcount\m**
\newcount\n*   \newcount\dn*  \newcount\r*
\newcount\N*   \newcount\*one \newcount\*two  \*one=1 \*two=2
\newcount\*ths \*ths=1000
\newcount\angle*  \newcount\q*  \newcount\q**
\newcount\angle** \angle**=0
\newcount\sc*     \sc*=0

\newtoks\cos*  \cos*={1}
\newtoks\sin*  \sin*={0}

\catcode`\[=13

\def\rotate(#1){\advance\angle**#1\angle*=\angle**
\q**=\angle*\ifnum\q**<0\q**=-\q**\fi
\ifnum\q**>360\q*=\angle*\divide\q*360\multiply\q*360\advance\angle*-\q*\fi
\ifnum\angle*<0\advance\angle*360\fi\q**=\angle*\divide\q**90\q**=\q**
\def\sgcos*{+}\def\sgsin*{+}\relax
\ifcase\q**\or
 \def\sgcos*{-}\def\sgsin*{+}\or
 \def\sgcos*{-}\def\sgsin*{-}\or
 \def\sgcos*{+}\def\sgsin*{-}\else\fi
\q*=\q**
\multiply\q*90\advance\angle*-\q*
\ifnum\angle*>45\sc*=1\angle*=-\angle*\advance\angle*90\else\sc*=0\fi
\def[##1,##2]{\ifnum\sc*=0\relax
\edef\cs*{\sgcos*.##1}\edef\sn*{\sgsin*.##2}\ifcase\q**\or
 \edef\cs*{\sgcos*.##2}\edef\sn*{\sgsin*.##1}\or
 \edef\cs*{\sgcos*.##1}\edef\sn*{\sgsin*.##2}\or
 \edef\cs*{\sgcos*.##2}\edef\sn*{\sgsin*.##1}\else\fi\else
\edef\cs*{\sgcos*.##2}\edef\sn*{\sgsin*.##1}\ifcase\q**\or
 \edef\cs*{\sgcos*.##1}\edef\sn*{\sgsin*.##2}\or
 \edef\cs*{\sgcos*.##2}\edef\sn*{\sgsin*.##1}\or
 \edef\cs*{\sgcos*.##1}\edef\sn*{\sgsin*.##2}\else\fi\fi
\cos*={\cs*}\sin*={\sn*}\global\edef\gcos*{\cs*}\global\edef\gsin*{\sn*}}\relax
\ifcase\angle*[9999,0]\or
[999,017]\or[999,034]\or[998,052]\or[997,069]\or[996,087]\or
[994,104]\or[992,121]\or[990,139]\or[987,156]\or[984,173]\or
[981,190]\or[978,207]\or[974,224]\or[970,241]\or[965,258]\or
[961,275]\or[956,292]\or[951,309]\or[945,325]\or[939,342]\or
[933,358]\or[927,374]\or[920,390]\or[913,406]\or[906,422]\or
[898,438]\or[891,453]\or[882,469]\or[874,484]\or[866,499]\or
[857,515]\or[848,529]\or[838,544]\or[829,559]\or[819,573]\or
[809,587]\or[798,601]\or[788,615]\or[777,629]\or[766,642]\or
[754,656]\or[743,669]\or[731,681]\or[719,694]\or[707,707]\or
\else[9999,0]\fi}

\catcode`\[=12

\def\GRAPH(hsize=#1)#2{\hbox to #1\Lengthunit{#2\hss}}

\def\Linewidth#1{\global\linwid*=#1\relax
\global\divide\linwid*10\global\multiply\linwid*\mag
\global\divide\linwid*100\special{em:linewidth \the\linwid*}}

\Linewidth{.4pt}
\def\sm*{\special{em:moveto}}
\def\sl*{\special{em:lineto}}
\let\moveto=\sm*
\let\lineto=\sl*
\newbox\spm*   \newbox\spl*
\setbox\spm*\hbox{\sm*}
\setbox\spl*\hbox{\sl*}

\def\mov#1(#2,#3)#4{\rlap{\L*=#1\Lengthunit
\xL*=#2\L* \yL*=#3\L*
\xL*=\xscale\xL* \yL*=\yscale\yL*
\rx* \the\cos*\xL* \tmp* \the\sin*\yL* \advance\rx*-\tmp*
\ry* \the\cos*\yL* \tmp* \the\sin*\xL* \advance\ry*\tmp*
\kern\rx*\raise\ry*\hbox{#4}}}

\def\rmov*(#1,#2)#3{\rlap{\xL*=#1\yL*=#2\relax
\rx* \the\cos*\xL* \tmp* \the\sin*\yL* \advance\rx*-\tmp*
\ry* \the\cos*\yL* \tmp* \the\sin*\xL* \advance\ry*\tmp*
\kern\rx*\raise\ry*\hbox{#3}}}

\def\lin#1(#2,#3){\rlap{\sm*\mov#1(#2,#3){\sl*}}}

\def\arr*(#1,#2,#3){\rmov*(#1\dd*,#1\dt*){\sm*
\rmov*(#2\dd*,#2\dt*){\rmov*(#3\dt*,-#3\dd*){\sl*}}\sm*
\rmov*(#2\dd*,#2\dt*){\rmov*(-#3\dt*,#3\dd*){\sl*}}}}

\def\arrow#1(#2,#3){\rlap{\lin#1(#2,#3)\mov#1(#2,#3){\relax
\d**=-.012\Lengthunit\dd*=#2\d**\dt*=#3\d**
\arr*(1,10,4)\arr*(3,8,4)\arr*(4.8,4.2,3)}}}

\def\arrlin#1(#2,#3){\rlap{\L*=#1\Lengthunit\L*=.5\L*
\lin#1(#2,#3)\rmov*(#2\L*,#3\L*){\arrow.1(#2,#3)}}}

\def\dasharrow#1(#2,#3){\rlap{{\Lengthunit=0.9\Lengthunit
\dashlin#1(#2,#3)\mov#1(#2,#3){\sm*}}\mov#1(#2,#3){\sl*
\d**=-.012\Lengthunit\dd*=#2\d**\dt*=#3\d**
\arr*(1,10,4)\arr*(3,8,4)\arr*(4.8,4.2,3)}}}

\def\clap#1{\hbox to 0pt{\hss #1\hss}}

\def\ind(#1,#2)#3{\rlap{\L*=.1\Lengthunit
\xL*=#1\L* \yL*=#2\L*
\rx* \the\cos*\xL* \tmp* \the\sin*\yL* \advance\rx*-\tmp*
\ry* \the\cos*\yL* \tmp* \the\sin*\xL* \advance\ry*\tmp*
\kern\rx*\raise\ry*\hbox{\lower2pt\clap{$#3$}}}}

\def\sh*(#1,#2)#3{\rlap{\dm*=\the\n*\d**
\xL*=\xscale\dm* \yL*=\yscale\dm* \xL*=#1\xL* \yL*=#2\yL*
\rx* \the\cos*\xL* \tmp* \the\sin*\yL* \advance\rx*-\tmp*
\ry* \the\cos*\yL* \tmp* \the\sin*\xL* \advance\ry*\tmp*
\kern\rx*\raise\ry*\hbox{#3}}}

\def\calcnum*#1(#2,#3){\a*=1000sp\b*=1000sp\a*=#2\a*\b*=#3\b*
\ifdim\a*<0pt\a*-\a*\fi\ifdim\b*<0pt\b*-\b*\fi
\ifdim\a*>\b*\c*=.96\a*\advance\c*.4\b*
\else\c*=.96\b*\advance\c*.4\a*\fi
\k*\a*\multiply\k*\k*\l*\b*\multiply\l*\l*
\m*\k*\advance\m*\l*\n*\c*\r*\n*\multiply\n*\n*
\dn*\m*\advance\dn*-\n*\divide\dn*2\divide\dn*\r*
\advance\r*\dn*
\c*=\the\Nhalfperiods5sp\c*=#1\c*\ifdim\c*<0pt\c*-\c*\fi
\multiply\c*\r*\N*\c*\divide\N*10000}

\def\dashlin#1(#2,#3){\rlap{\calcnum*#1(#2,#3)\relax
\d**=#1\Lengthunit\ifdim\d**<0pt\d**-\d**\fi
\divide\N*2\multiply\N*2\advance\N*\*one
\divide\d**\N*\sm*\n*\*one\sh*(#2,#3){\sl*}\loop
\advance\n*\*one\sh*(#2,#3){\sm*}\advance\n*\*one
\sh*(#2,#3){\sl*}\ifnum\n*<\N*\repeat}}

\def\dashdotlin#1(#2,#3){\rlap{\calcnum*#1(#2,#3)\relax
\d**=#1\Lengthunit\ifdim\d**<0pt\d**-\d**\fi
\divide\N*2\multiply\N*2\advance\N*1\multiply\N*2\relax
\divide\d**\N*\sm*\n*\*two\sh*(#2,#3){\sl*}\loop
\advance\n*\*one\sh*(#2,#3){\kern-1.48pt\lower.5pt\hbox{\rm.}}\relax
\advance\n*\*one\sh*(#2,#3){\sm*}\advance\n*\*two
\sh*(#2,#3){\sl*}\ifnum\n*<\N*\repeat}}

\def\shl*(#1,#2)#3{\kern#1#3\lower#2#3\hbox{\unhcopy\spl*}}

\def\trianglin#1(#2,#3){\rlap{\toks0={#2}\toks1={#3}\calcnum*#1(#2,#3)\relax
\dd*=.57\Lengthunit\dd*=#1\dd*\divide\dd*\N*
\divide\dd*\*ths \multiply\dd*\magnitude
\d**=#1\Lengthunit\ifdim\d**<0pt\d**-\d**\fi
\multiply\N*2\divide\d**\N*\sm*\n*\*one\loop
\shl**{\dd*}\dd*-\dd*\advance\n*2\relax
\ifnum\n*<\N*\repeat\n*\N*\shl**{0pt}}}

\def\wavelin#1(#2,#3){\rlap{\toks0={#2}\toks1={#3}\calcnum*#1(#2,#3)\relax
\dd*=.23\Lengthunit\dd*=#1\dd*\divide\dd*\N*
\divide\dd*\*ths \multiply\dd*\magnitude
\d**=#1\Lengthunit\ifdim\d**<0pt\d**-\d**\fi
\multiply\N*4\divide\d**\N*\sm*\n*\*one\loop
\shl**{\dd*}\dt*=1.3\dd*\advance\n*\*one
\shl**{\dt*}\advance\n*\*one
\shl**{\dd*}\advance\n*\*two
\dd*-\dd*\ifnum\n*<\N*\repeat\n*\N*\shl**{0pt}}}

\def\w*lin(#1,#2){\rlap{\toks0={#1}\toks1={#2}\d**=\Lengthunit\dd*=-.12\d**
\divide\dd*\*ths \multiply\dd*\magnitude
\N*8\divide\d**\N*\sm*\n*\*one\loop
\shl**{\dd*}\dt*=1.3\dd*\advance\n*\*one
\shl**{\dt*}\advance\n*\*one
\shl**{\dd*}\advance\n*\*one
\shl**{0pt}\dd*-\dd*\advance\n*1\ifnum\n*<\N*\repeat}}

\def\l*arc(#1,#2)[#3][#4]{\rlap{\toks0={#1}\toks1={#2}\d**=\Lengthunit
\dd*=#3.037\d**\dd*=#4\dd*\dt*=#3.049\d**\dt*=#4\dt*\ifdim\d**>10mm\relax
\d**=.25\d**\n*\*one\shl**{-\dd*}\n*\*two\shl**{-\dt*}\n*3\relax
\shl**{-\dd*}\n*4\relax\shl**{0pt}\else
\ifdim\d**>5mm\d**=.5\d**\n*\*one\shl**{-\dt*}\n*\*two
\shl**{0pt}\else\n*\*one\shl**{0pt}\fi\fi}}

\def\d*arc(#1,#2)[#3][#4]{\rlap{\toks0={#1}\toks1={#2}\d**=\Lengthunit
\dd*=#3.037\d**\dd*=#4\dd*\d**=.25\d**\sm*\n*\*one\shl**{-\dd*}\relax
\n*3\relax\sh*(#1,#2){\xL*=\xscale\dd*\yL*=\yscale\dd*
\kern#2\xL*\lower#1\yL*\hbox{\sm*}}\n*4\relax\shl**{0pt}}}

\def\shl**#1{\c*=\the\n*\d**\d*=#1\relax
\a*=\the\toks0\c*\b*=\the\toks1\d*\advance\a*-\b*
\b*=\the\toks1\c*\d*=\the\toks0\d*\advance\b*\d*
\a*=\xscale\a*\b*=\yscale\b*
\rx* \the\cos*\a* \tmp* \the\sin*\b* \advance\rx*-\tmp*
\ry* \the\cos*\b* \tmp* \the\sin*\a* \advance\ry*\tmp*
\raise\ry*\rlap{\kern\rx*\unhcopy\spl*}}

\def\wlin*#1(#2,#3)[#4]{\rlap{\toks0={#2}\toks1={#3}\relax
\c*=#1\l*\c*\c*=.01\Lengthunit\m*\c*\divide\l*\m*
\c*=\the\Nhalfperiods5sp\multiply\c*\l*\N*\c*\divide\N*\*ths
\divide\N*2\multiply\N*2\advance\N*\*one
\dd*=.002\Lengthunit\dd*=#4\dd*\multiply\dd*\l*\divide\dd*\N*
\divide\dd*\*ths \multiply\dd*\magnitude
\d**=#1\multiply\N*4\divide\d**\N*\sm*\n*\*one\loop
\shl**{\dd*}\dt*=1.3\dd*\advance\n*\*one
\shl**{\dt*}\advance\n*\*one
\shl**{\dd*}\advance\n*\*two
\dd*-\dd*\ifnum\n*<\N*\repeat\n*\N*\shl**{0pt}}}

\def\wavebox#1{\setbox0\hbox{#1}\relax
\a*=\wd0\advance\a*14pt\b*=\ht0\advance\b*\dp0\advance\b*14pt\relax
\hbox{\kern9pt\relax
\rmov*(0pt,\ht0){\rmov*(-7pt,7pt){\wlin*\a*(1,0)[+]\wlin*\b*(0,-1)[-]}}\relax
\rmov*(\wd0,-\dp0){\rmov*(7pt,-7pt){\wlin*\a*(-1,0)[+]\wlin*\b*(0,1)[-]}}\relax
\box0\kern9pt}}

\def\rectangle#1(#2,#3){\relax
\lin#1(#2,0)\lin#1(0,#3)\mov#1(0,#3){\lin#1(#2,0)}\mov#1(#2,0){\lin#1(0,#3)}}

\def\dashrectangle#1(#2,#3){\dashlin#1(#2,0)\dashlin#1(0,#3)\relax
\mov#1(0,#3){\dashlin#1(#2,0)}\mov#1(#2,0){\dashlin#1(0,#3)}}

\def\waverectangle#1(#2,#3){\L*=#1\Lengthunit\a*=#2\L*\b*=#3\L*
\ifdim\a*<0pt\a*-\a*\def\x*{-1}\else\def\x*{1}\fi
\ifdim\b*<0pt\b*-\b*\def\y*{-1}\else\def\y*{1}\fi
\wlin*\a*(\x*,0)[-]\wlin*\b*(0,\y*)[+]\relax
\mov#1(0,#3){\wlin*\a*(\x*,0)[+]}\mov#1(#2,0){\wlin*\b*(0,\y*)[-]}}

\def\calcparab*{\ifnum\n*>\m*\k*\N*\advance\k*-\n*\else\k*\n*\fi
\a*=\the\k* sp\a*=10\a*\b*\dm*\advance\b*-\a*\k*\b*
\a*=\the\*ths\b*\divide\a*\l*\multiply\a*\k*
\divide\a*\l*\k*\*ths\r*\a*\advance\k*-\r*\dt*=\the\k*\L*}

\def\arcto#1(#2,#3)[#4]{\rlap{\toks0={#2}\toks1={#3}\calcnum*#1(#2,#3)\relax
\dm*=135sp\dm*=#1\dm*\d**=#1\Lengthunit\ifdim\dm*<0pt\dm*-\dm*\fi
\multiply\dm*\r*\a*=.3\dm*\a*=#4\a*\ifdim\a*<0pt\a*-\a*\fi
\advance\dm*\a*\N*\dm*\divide\N*10000\relax
\divide\N*2\multiply\N*2\advance\N*\*one
\L*=-.25\d**\L*=#4\L*\divide\d**\N*\divide\L*\*ths
\m*\N*\divide\m*2\dm*=\the\m*5sp\l*\dm*\sm*\n*\*one\loop
\calcparab*\shl**{-\dt*}\advance\n*1\ifnum\n*<\N*\repeat}}

\def\arrarcto#1(#2,#3)[#4]{\L*=#1\Lengthunit\L*=.54\L*
\arcto#1(#2,#3)[#4]\rmov*(#2\L*,#3\L*){\d*=.457\L*\d*=#4\d*\d**-\d*
\rmov*(#3\d**,#2\d*){\arrow.02(#2,#3)}}}

\def\dasharcto#1(#2,#3)[#4]{\rlap{\toks0={#2}\toks1={#3}\relax
\calcnum*#1(#2,#3)\dm*=\the\N*5sp\a*=.3\dm*\a*=#4\a*\ifdim\a*<0pt\a*-\a*\fi
\advance\dm*\a*\N*\dm*
\divide\N*20\multiply\N*2\advance\N*1\d**=#1\Lengthunit
\L*=-.25\d**\L*=#4\L*\divide\d**\N*\divide\L*\*ths
\m*\N*\divide\m*2\dm*=\the\m*5sp\l*\dm*
\sm*\n*\*one\loop\calcparab*
\shl**{-\dt*}\advance\n*1\ifnum\n*>\N*\else\calcparab*
\sh*(#2,#3){\xL*=#3\dt* \yL*=#2\dt*
\rx* \the\cos*\xL* \tmp* \the\sin*\yL* \advance\rx*\tmp*
\ry* \the\cos*\yL* \tmp* \the\sin*\xL* \advance\ry*-\tmp*
\kern\rx*\lower\ry*\hbox{\sm*}}\fi
\advance\n*1\ifnum\n*<\N*\repeat}}

\def\*shl*#1{\c*=\the\n*\d**\advance\c*#1\a**\d*\dt*\advance\d*#1\b**
\a*=\the\toks0\c*\b*=\the\toks1\d*\advance\a*-\b*
\b*=\the\toks1\c*\d*=\the\toks0\d*\advance\b*\d*
\rx* \the\cos*\a* \tmp* \the\sin*\b* \advance\rx*-\tmp*
\ry* \the\cos*\b* \tmp* \the\sin*\a* \advance\ry*\tmp*
\raise\ry*\rlap{\kern\rx*\unhcopy\spl*}}

\def\calcnormal*#1{\b**=10000sp\a**\b**\k*\n*\advance\k*-\m*
\multiply\a**\k*\divide\a**\m*\a**=#1\a**\ifdim\a**<0pt\a**-\a**\fi
\ifdim\a**>\b**\d*=.96\a**\advance\d*.4\b**
\else\d*=.96\b**\advance\d*.4\a**\fi
\d*=.01\d*\r*\d*\divide\a**\r*\divide\b**\r*
\ifnum\k*<0\a**-\a**\fi\d*=#1\d*\ifdim\d*<0pt\b**-\b**\fi
\k*\a**\a**=\the\k*\dd*\k*\b**\b**=\the\k*\dd*}

\def\wavearcto#1(#2,#3)[#4]{\rlap{\toks0={#2}\toks1={#3}\relax
\calcnum*#1(#2,#3)\c*=\the\N*5sp\a*=.4\c*\a*=#4\a*\ifdim\a*<0pt\a*-\a*\fi
\advance\c*\a*\N*\c*\divide\N*20\multiply\N*2\advance\N*-1\multiply\N*4\relax
\d**=#1\Lengthunit\dd*=.012\d**
\divide\dd*\*ths \multiply\dd*\magnitude
\ifdim\d**<0pt\d**-\d**\fi\L*=.25\d**
\divide\d**\N*\divide\dd*\N*\L*=#4\L*\divide\L*\*ths
\m*\N*\divide\m*2\dm*=\the\m*0sp\l*\dm*
\sm*\n*\*one\loop\calcnormal*{#4}\calcparab*
\*shl*{1}\advance\n*\*one\calcparab*
\*shl*{1.3}\advance\n*\*one\calcparab*
\*shl*{1}\advance\n*2\dd*-\dd*\ifnum\n*<\N*\repeat\n*\N*\shl**{0pt}}}

\def\triangarcto#1(#2,#3)[#4]{\rlap{\toks0={#2}\toks1={#3}\relax
\calcnum*#1(#2,#3)\c*=\the\N*5sp\a*=.4\c*\a*=#4\a*\ifdim\a*<0pt\a*-\a*\fi
\advance\c*\a*\N*\c*\divide\N*20\multiply\N*2\advance\N*-1\multiply\N*2\relax
\d**=#1\Lengthunit\dd*=.012\d**
\divide\dd*\*ths \multiply\dd*\magnitude
\ifdim\d**<0pt\d**-\d**\fi\L*=.25\d**
\divide\d**\N*\divide\dd*\N*\L*=#4\L*\divide\L*\*ths
\m*\N*\divide\m*2\dm*=\the\m*0sp\l*\dm*
\sm*\n*\*one\loop\calcnormal*{#4}\calcparab*
\*shl*{1}\advance\n*2\dd*-\dd*\ifnum\n*<\N*\repeat\n*\N*\shl**{0pt}}}

\def\hr*#1{\L*=\xscale\Lengthunit\ifnum
\angle**=0\clap{\vrule width#1\L* height.1pt}\else
\L*=#1\L*\L*=.5\L*\rmov*(-\L*,0pt){\sm*}\rmov*(\L*,0pt){\sl*}\fi}

\def\shade#1[#2]{\rlap{\Lengthunit=#1\Lengthunit
\special{em:linewidth .001pt}\relax
\mov(0,#2.05){\hr*{.994}}\mov(0,#2.1){\hr*{.980}}\relax
\mov(0,#2.15){\hr*{.953}}\mov(0,#2.2){\hr*{.916}}\relax
\mov(0,#2.25){\hr*{.867}}\mov(0,#2.3){\hr*{.798}}\relax
\mov(0,#2.35){\hr*{.715}}\mov(0,#2.4){\hr*{.603}}\relax
\mov(0,#2.45){\hr*{.435}}\special{em:linewidth \the\linwid*}}}

\def\dshade#1[#2]{\rlap{\special{em:linewidth .001pt}\relax
\Lengthunit=#1\Lengthunit\if#2-\def\t*{+}\else\def\t*{-}\fi
\mov(0,\t*.025){\relax
\mov(0,#2.05){\hr*{.995}}\mov(0,#2.1){\hr*{.988}}\relax
\mov(0,#2.15){\hr*{.969}}\mov(0,#2.2){\hr*{.937}}\relax
\mov(0,#2.25){\hr*{.893}}\mov(0,#2.3){\hr*{.836}}\relax
\mov(0,#2.35){\hr*{.760}}\mov(0,#2.4){\hr*{.662}}\relax
\mov(0,#2.45){\hr*{.531}}\mov(0,#2.5){\hr*{.320}}\relax
\special{em:linewidth \the\linwid*}}}}

\def\vdot{\rlap{\kern-1.9pt\lower1.8pt\hbox{$\scriptstyle\bullet$}}}
\def\vtimes{\rlap{\kern-3pt\lower1.8pt\hbox{$\scriptstyle\times$}}}
\def\vDot{\rlap{\kern-2.3pt\lower2.7pt\hbox{$\bullet$}}}
\def\vTimes{\rlap{\kern-3.6pt\lower2.4pt\hbox{$\times$}}}

\def\arc(#1)[#2,#3]{{\k*=#2\l*=#3\m*=\l*
\advance\m*-6\ifnum\k*>\l*\relax\else
{\rotate(#2)\mov(#1,0){\sm*}}\loop
\ifnum\k*<\m*\advance\k*5{\rotate(\k*)\mov(#1,0){\sl*}}\repeat
{\rotate(#3)\mov(#1,0){\sl*}}\fi}}

\def\dasharc(#1)[#2,#3]{{\k**=#2\n*=#3\advance\n*-1\advance\n*-\k**
\L*=1000sp\L*#1\L* \multiply\L*\n* \multiply\L*\Nhalfperiods
\divide\L*57\N*\L* \divide\N*2000\ifnum\N*=0\N*1\fi
\r*\n*  \divide\r*\N* \ifnum\r*<2\r*2\fi
\m**\r* \divide\m**2 \l**\r* \advance\l**-\m** \N*\n* \divide\N*\r*
\k**\r* \multiply\k**\N* \dn*\n*
\advance\dn*-\k** \divide\dn*2\advance\dn*\*one
\r*\l** \divide\r*2\advance\dn*\r* \advance\N*-2\k**#2\relax
\ifnum\l**<6{\rotate(#2)\mov(#1,0){\sm*}}\advance\k**\dn*
{\rotate(\k**)\mov(#1,0){\sl*}}\advance\k**\m**
{\rotate(\k**)\mov(#1,0){\sm*}}\loop
\advance\k**\l**{\rotate(\k**)\mov(#1,0){\sl*}}\advance\k**\m**
{\rotate(\k**)\mov(#1,0){\sm*}}\advance\N*-1\ifnum\N*>0\repeat
{\rotate(#3)\mov(#1,0){\sl*}}\else\advance\k**\dn*
\arc(#1)[#2,\k**]\loop\advance\k**\m** \r*\k**
\advance\k**\l** {\arc(#1)[\r*,\k**]}\relax
\advance\N*-1\ifnum\N*>0\repeat
\advance\k**\m**\arc(#1)[\k**,#3]\fi}}

\def\triangarc#1(#2)[#3,#4]{{\k**=#3\n*=#4\advance\n*-\k**
\L*=1000sp\L*#2\L* \multiply\L*\n* \multiply\L*\Nhalfperiods
\divide\L*57\N*\L* \divide\N*1000\ifnum\N*=0\N*1\fi
\d**=#2\Lengthunit \d*\d** \divide\d*57\multiply\d*\n*
\r*\n*  \divide\r*\N* \ifnum\r*<2\r*2\fi
\m**\r* \divide\m**2 \l**\r* \advance\l**-\m** \N*\n* \divide\N*\r*
\dt*\d* \divide\dt*\N* \dt*.5\dt* \dt*#1\dt*
\divide\dt*1000\multiply\dt*\magnitude
\k**\r* \multiply\k**\N* \dn*\n* \advance\dn*-\k** \divide\dn*2\relax
\r*\l** \divide\r*2\advance\dn*\r* \advance\N*-1\k**#3\relax
{\rotate(#3)\mov(#2,0){\sm*}}\advance\k**\dn*
{\rotate(\k**)\mov(#2,0){\sl*}}\advance\k**-\m**\advance\l**\m**\loop\dt*-\dt*
\d*\d** \advance\d*\dt*
\advance\k**\l**{\rotate(\k**)\rmov*(\d*,0pt){\sl*}}%
\advance\N*-1\ifnum\N*>0\repeat\advance\k**\m**
{\rotate(\k**)\mov(#2,0){\sl*}}{\rotate(#4)\mov(#2,0){\sl*}}}}

\def\wavearc#1(#2)[#3,#4]{{\k**=#3\n*=#4\advance\n*-\k**
\L*=4000sp\L*#2\L* \multiply\L*\n* \multiply\L*\Nhalfperiods
\divide\L*57\N*\L* \divide\N*1000\ifnum\N*=0\N*1\fi
\d**=#2\Lengthunit \d*\d** \divide\d*57\multiply\d*\n*
\r*\n*  \divide\r*\N* \ifnum\r*=0\r*1\fi
\m**\r* \divide\m**2 \l**\r* \advance\l**-\m** \N*\n* \divide\N*\r*
\dt*\d* \divide\dt*\N* \dt*.7\dt* \dt*#1\dt*
\divide\dt*1000\multiply\dt*\magnitude
\k**\r* \multiply\k**\N* \dn*\n* \advance\dn*-\k** \divide\dn*2\relax
\divide\N*4\advance\N*-1\k**#3\relax
{\rotate(#3)\mov(#2,0){\sm*}}\advance\k**\dn*
{\rotate(\k**)\mov(#2,0){\sl*}}\advance\k**-\m**\advance\l**\m**\loop\dt*-\dt*
\d*\d** \advance\d*\dt* \dd*\d** \advance\dd*1.3\dt*
\advance\k**\r*{\rotate(\k**)\rmov*(\d*,0pt){\sl*}}\relax
\advance\k**\r*{\rotate(\k**)\rmov*(\dd*,0pt){\sl*}}\relax
\advance\k**\r*{\rotate(\k**)\rmov*(\d*,0pt){\sl*}}\relax
\advance\k**\r*
\advance\N*-1\ifnum\N*>0\repeat\advance\k**\m**
{\rotate(\k**)\mov(#2,0){\sl*}}{\rotate(#4)\mov(#2,0){\sl*}}}}

\def\gmov*#1(#2,#3)#4{\rlap{\L*=#1\Lengthunit
\xL*=#2\L* \yL*=#3\L*
\rx* \gcos*\xL* \tmp* \gsin*\yL* \advance\rx*-\tmp*
\ry* \gcos*\yL* \tmp* \gsin*\xL* \advance\ry*\tmp*
\rx*=\xscale\rx* \ry*=\yscale\ry*
\xL* \the\cos*\rx* \tmp* \the\sin*\ry* \advance\xL*-\tmp*
\yL* \the\cos*\ry* \tmp* \the\sin*\rx* \advance\yL*\tmp*
\kern\xL*\raise\yL*\hbox{#4}}}

\def\rgmov*(#1,#2)#3{\rlap{\xL*#1\yL*#2\relax
\rx* \gcos*\xL* \tmp* \gsin*\yL* \advance\rx*-\tmp*
\ry* \gcos*\yL* \tmp* \gsin*\xL* \advance\ry*\tmp*
\rx*=\xscale\rx* \ry*=\yscale\ry*
\xL* \the\cos*\rx* \tmp* \the\sin*\ry* \advance\xL*-\tmp*
\yL* \the\cos*\ry* \tmp* \the\sin*\rx* \advance\yL*\tmp*
\kern\xL*\raise\yL*\hbox{#3}}}

\def\Earc(#1)[#2,#3][#4,#5]{{\k*=#2\l*=#3\m*=\l*
\advance\m*-6\ifnum\k*>\l*\relax\else\def\xscale{#4}\def\yscale{#5}\relax
{\angle**0\rotate(#2)}\gmov*(#1,0){\sm*}\loop
\ifnum\k*<\m*\advance\k*5\relax
{\angle**0\rotate(\k*)}\gmov*(#1,0){\sl*}\repeat
{\angle**0\rotate(#3)}\gmov*(#1,0){\sl*}\relax
\def\xscale{1}\def\yscale{1}\fi}}

\def\dashEarc(#1)[#2,#3][#4,#5]{{\k**=#2\n*=#3\advance\n*-1\advance\n*-\k**
\L*=1000sp\L*#1\L* \multiply\L*\n* \multiply\L*\Nhalfperiods
\divide\L*57\N*\L* \divide\N*2000\ifnum\N*=0\N*1\fi
\r*\n*  \divide\r*\N* \ifnum\r*<2\r*2\fi
\m**\r* \divide\m**2 \l**\r* \advance\l**-\m** \N*\n* \divide\N*\r*
\k**\r*\multiply\k**\N* \dn*\n* \advance\dn*-\k** \divide\dn*2\advance\dn*\*one
\r*\l** \divide\r*2\advance\dn*\r* \advance\N*-2\k**#2\relax
\ifnum\l**<6\def\xscale{#4}\def\yscale{#5}\relax
{\angle**0\rotate(#2)}\gmov*(#1,0){\sm*}\advance\k**\dn*
{\angle**0\rotate(\k**)}\gmov*(#1,0){\sl*}\advance\k**\m**
{\angle**0\rotate(\k**)}\gmov*(#1,0){\sm*}\loop
\advance\k**\l**{\angle**0\rotate(\k**)}\gmov*(#1,0){\sl*}\advance\k**\m**
{\angle**0\rotate(\k**)}\gmov*(#1,0){\sm*}\advance\N*-1\ifnum\N*>0\repeat
{\angle**0\rotate(#3)}\gmov*(#1,0){\sl*}\def\xscale{1}\def\yscale{1}\else
\advance\k**\dn* \Earc(#1)[#2,\k**][#4,#5]\loop\advance\k**\m** \r*\k**
\advance\k**\l** {\Earc(#1)[\r*,\k**][#4,#5]}\relax
\advance\N*-1\ifnum\N*>0\repeat
\advance\k**\m**\Earc(#1)[\k**,#3][#4,#5]\fi}}

\def\triangEarc#1(#2)[#3,#4][#5,#6]{{\k**=#3\n*=#4\advance\n*-\k**
\L*=1000sp\L*#2\L* \multiply\L*\n* \multiply\L*\Nhalfperiods
\divide\L*57\N*\L* \divide\N*1000\ifnum\N*=0\N*1\fi
\d**=#2\Lengthunit \d*\d** \divide\d*57\multiply\d*\n*
\r*\n*  \divide\r*\N* \ifnum\r*<2\r*2\fi
\m**\r* \divide\m**2 \l**\r* \advance\l**-\m** \N*\n* \divide\N*\r*
\dt*\d* \divide\dt*\N* \dt*.5\dt* \dt*#1\dt*
\divide\dt*1000\multiply\dt*\magnitude
\k**\r* \multiply\k**\N* \dn*\n* \advance\dn*-\k** \divide\dn*2\relax
\r*\l** \divide\r*2\advance\dn*\r* \advance\N*-1\k**#3\relax
\def\xscale{#5}\def\yscale{#6}\relax
{\angle**0\rotate(#3)}\gmov*(#2,0){\sm*}\advance\k**\dn*
{\angle**0\rotate(\k**)}\gmov*(#2,0){\sl*}\advance\k**-\m**
\advance\l**\m**\loop\dt*-\dt* \d*\d** \advance\d*\dt*
\advance\k**\l**{\angle**0\rotate(\k**)}\rgmov*(\d*,0pt){\sl*}\relax
\advance\N*-1\ifnum\N*>0\repeat\advance\k**\m**
{\angle**0\rotate(\k**)}\gmov*(#2,0){\sl*}\relax
{\angle**0\rotate(#4)}\gmov*(#2,0){\sl*}\def\xscale{1}\def\yscale{1}}}

\def\waveEarc#1(#2)[#3,#4][#5,#6]{{\k**=#3\n*=#4\advance\n*-\k**
\L*=4000sp\L*#2\L* \multiply\L*\n* \multiply\L*\Nhalfperiods
\divide\L*57\N*\L* \divide\N*1000\ifnum\N*=0\N*1\fi
\d**=#2\Lengthunit \d*\d** \divide\d*57\multiply\d*\n*
\r*\n*  \divide\r*\N* \ifnum\r*=0\r*1\fi
\m**\r* \divide\m**2 \l**\r* \advance\l**-\m** \N*\n* \divide\N*\r*
\dt*\d* \divide\dt*\N* \dt*.7\dt* \dt*#1\dt*
\divide\dt*1000\multiply\dt*\magnitude
\k**\r* \multiply\k**\N* \dn*\n* \advance\dn*-\k** \divide\dn*2\relax
\divide\N*4\advance\N*-1\k**#3\def\xscale{#5}\def\yscale{#6}\relax
{\angle**0\rotate(#3)}\gmov*(#2,0){\sm*}\advance\k**\dn*
{\angle**0\rotate(\k**)}\gmov*(#2,0){\sl*}\advance\k**-\m**
\advance\l**\m**\loop\dt*-\dt*
\d*\d** \advance\d*\dt* \dd*\d** \advance\dd*1.3\dt*
\advance\k**\r*{\angle**0\rotate(\k**)}\rgmov*(\d*,0pt){\sl*}\relax
\advance\k**\r*{\angle**0\rotate(\k**)}\rgmov*(\dd*,0pt){\sl*}\relax
\advance\k**\r*{\angle**0\rotate(\k**)}\rgmov*(\d*,0pt){\sl*}\relax
\advance\k**\r*
\advance\N*-1\ifnum\N*>0\repeat\advance\k**\m**
{\angle**0\rotate(\k**)}\gmov*(#2,0){\sl*}\relax
{\angle**0\rotate(#4)}\gmov*(#2,0){\sl*}\def\xscale{1}\def\yscale{1}}}

\newcount\CatcodeOfAtSign
\CatcodeOfAtSign=\the\catcode`\@
\catcode`\@=11
\def\@arc#1[#2][#3]{\rlap{\Lengthunit=#1\Lengthunit
\sm*\l*arc(#2.1914,#3.0381)[#2][#3]\relax
\mov(#2.1914,#3.0381){\l*arc(#2.1622,#3.1084)[#2][#3]}\relax
\mov(#2.3536,#3.1465){\l*arc(#2.1084,#3.1622)[#2][#3]}\relax
\mov(#2.4619,#3.3086){\l*arc(#2.0381,#3.1914)[#2][#3]}}}

\def\dash@arc#1[#2][#3]{\rlap{\Lengthunit=#1\Lengthunit
\d*arc(#2.1914,#3.0381)[#2][#3]\relax
\mov(#2.1914,#3.0381){\d*arc(#2.1622,#3.1084)[#2][#3]}\relax
\mov(#2.3536,#3.1465){\d*arc(#2.1084,#3.1622)[#2][#3]}\relax
\mov(#2.4619,#3.3086){\d*arc(#2.0381,#3.1914)[#2][#3]}}}

\def\wave@arc#1[#2][#3]{\rlap{\Lengthunit=#1\Lengthunit
\w*lin(#2.1914,#3.0381)\relax
\mov(#2.1914,#3.0381){\w*lin(#2.1622,#3.1084)}\relax
\mov(#2.3536,#3.1465){\w*lin(#2.1084,#3.1622)}\relax
\mov(#2.4619,#3.3086){\w*lin(#2.0381,#3.1914)}}}

\def\bezier#1(#2,#3)(#4,#5)(#6,#7){\N*#1\l*\N* \advance\l*\*one
\d* #4\Lengthunit \advance\d* -#2\Lengthunit \multiply\d* \*two
\b* #6\Lengthunit \advance\b* -#2\Lengthunit
\advance\b*-\d* \divide\b*\N*
\d** #5\Lengthunit \advance\d** -#3\Lengthunit \multiply\d** \*two
\b** #7\Lengthunit \advance\b** -#3\Lengthunit
\advance\b** -\d** \divide\b**\N*
\mov(#2,#3){\sm*{\loop\ifnum\m*<\l*
\a*\m*\b* \advance\a*\d* \divide\a*\N* \multiply\a*\m*
\a**\m*\b** \advance\a**\d** \divide\a**\N* \multiply\a**\m*
\rmov*(\a*,\a**){\unhcopy\spl*}\advance\m*\*one\repeat}}}

\catcode`\*=12

\newcount\n@ast

\def\n@ast@#1{\n@ast0\relax\get@ast@#1\end}
\def\get@ast@#1{\ifx#1\end\let\next\relax\else
\ifx#1*\advance\n@ast1\fi\let\next\get@ast@\fi\next}

\newif\if@up \newif\if@dwn
\def\up@down@#1{\@upfalse\@dwnfalse
\if#1u\@uptrue\fi\if#1U\@uptrue\fi\if#1+\@uptrue\fi
\if#1d\@dwntrue\fi\if#1D\@dwntrue\fi\if#1-\@dwntrue\fi}

\def\halfcirc#1(#2)[#3]{{\Lengthunit=#2\Lengthunit\up@down@{#3}\relax
\if@up\mov(0,.5){\@arc[-][-]\@arc[+][-]}\fi
\if@dwn\mov(0,-.5){\@arc[-][+]\@arc[+][+]}\fi
\def\lft{\mov(0,.5){\@arc[-][-]}\mov(0,-.5){\@arc[-][+]}}\relax
\def\rght{\mov(0,.5){\@arc[+][-]}\mov(0,-.5){\@arc[+][+]}}\relax
\if#3l\lft\fi\if#3L\lft\fi\if#3r\rght\fi\if#3R\rght\fi
\n@ast@{#1}\relax
\ifnum\n@ast>0\if@up\shade[+]\fi\if@dwn\shade[-]\fi\fi
\ifnum\n@ast>1\if@up\dshade[+]\fi\if@dwn\dshade[-]\fi\fi}}

\def\halfdashcirc(#1)[#2]{{\Lengthunit=#1\Lengthunit\up@down@{#2}\relax
\if@up\mov(0,.5){\dash@arc[-][-]\dash@arc[+][-]}\fi
\if@dwn\mov(0,-.5){\dash@arc[-][+]\dash@arc[+][+]}\fi
\def\lft{\mov(0,.5){\dash@arc[-][-]}\mov(0,-.5){\dash@arc[-][+]}}\relax
\def\rght{\mov(0,.5){\dash@arc[+][-]}\mov(0,-.5){\dash@arc[+][+]}}\relax
\if#2l\lft\fi\if#2L\lft\fi\if#2r\rght\fi\if#2R\rght\fi}}

\def\halfwavecirc(#1)[#2]{{\Lengthunit=#1\Lengthunit\up@down@{#2}\relax
\if@up\mov(0,.5){\wave@arc[-][-]\wave@arc[+][-]}\fi
\if@dwn\mov(0,-.5){\wave@arc[-][+]\wave@arc[+][+]}\fi
\def\lft{\mov(0,.5){\wave@arc[-][-]}\mov(0,-.5){\wave@arc[-][+]}}\relax
\def\rght{\mov(0,.5){\wave@arc[+][-]}\mov(0,-.5){\wave@arc[+][+]}}\relax
\if#2l\lft\fi\if#2L\lft\fi\if#2r\rght\fi\if#2R\rght\fi}}

\catcode`\*=11

\def\Circle#1(#2){\halfcirc#1(#2)[u]\halfcirc#1(#2)[d]\n@ast@{#1}\relax
\ifnum\n@ast>0\L*=\xscale\Lengthunit
\ifnum\angle**=0\clap{\vrule width#2\L* height.1pt}\else
\L*=#2\L*\L*=.5\L*\special{em:linewidth .001pt}\relax
\rmov*(-\L*,0pt){\sm*}\rmov*(\L*,0pt){\sl*}\relax
\special{em:linewidth \the\linwid*}\fi\fi}

\catcode`\*=12

\def\wavecirc(#1){\halfwavecirc(#1)[u]\halfwavecirc(#1)[d]}
\def\dashcirc(#1){\halfdashcirc(#1)[u]\halfdashcirc(#1)[d]}

\def\xscale{1}

\def\yscale{1}

\def\Ellipse#1(#2)[#3,#4]{\def\xscale{#3}\def\yscale{#4}\relax
\Circle#1(#2)\def\xscale{1}\def\yscale{1}}

\def\dashEllipse(#1)[#2,#3]{\def\xscale{#2}\def\yscale{#3}\relax
\dashcirc(#1)\def\xscale{1}\def\yscale{1}}

\def\waveEllipse(#1)[#2,#3]{\def\xscale{#2}\def\yscale{#3}\relax
\wavecirc(#1)\def\xscale{1}\def\yscale{1}}

\def\halfEllipse#1(#2)[#3][#4,#5]{\def\xscale{#4}\def\yscale{#5}\relax
\halfcirc#1(#2)[#3]\def\xscale{1}\def\yscale{1}}

\def\halfdashEllipse(#1)[#2][#3,#4]{\def\xscale{#3}\def\yscale{#4}\relax
\halfdashcirc(#1)[#2]\def\xscale{1}\def\yscale{1}}

\def\halfwaveEllipse(#1)[#2][#3,#4]{\def\xscale{#3}\def\yscale{#4}\relax
\halfwavecirc(#1)[#2]\def\xscale{1}\def\yscale{1}}

\catcode`\@=\the\CatcodeOfAtSign

\title{\boldmath Baby Skyrme model and fermionic zero modes}

\author{J. M. Queiruga}
\affiliation{Institute for Theoretical Physics, Karlsruhe Institute
of Technology (KIT), 76131 Karlsruhe, Germany}
\email{jose.queiruga@kit.edu}
\preprint{KA-TP-17-2016}

\begin{abstract}
In this work we investigate some features of the fermionic sector of the supersymmetric version of the baby Skyrme model. We find that, in the background of BPS compact baby Skyrmions, fermionic zero modes are confined to the defect core. Further, we show that, while three supersymmetric (SUSY) generators are broken in the defect core, SUSY is completely restored outside. We study also the effect of a D-term deformation of the model. Such a deformation allows for the existence of fermionic zero modes and broken SUSY outside the compact defect.
\end{abstract}

\maketitle

\section{Introduction}
The Skyrme model, originally proposed by T. M. Skyrme \cite{Skyrme1,Skyrme2}, is a nonlinear field theory in four dimensional Minkowski space. It is one of the most well-known proposals for the study of the low energy nonperturbative QCD. Its physical degrees of freedom are described by  fields taking values in $SU(2)$. In the static case they are maps from the one-point compactification of the three dimensional space $\mathbb{R}^3\cup \{p\}$ to $SU(2)$. Since $\mathbb{R}^3\cup \{p\}\simeq \mathbb{S}^3\simeq SU(2)$, the field configurations can be classified by the degree of a map $\varphi: \mathbb{S}^3\rightarrow  \mathbb{S}^3$, which is an integer value called topological degree or winding number.

The low dimensional analogue of the Skyrme model is the so-called baby Skyrme model (bS) (or planar Skyrme model), \cite{Piette1,Piette2,Leese,Sutcliffe, Weidig, Gisiger}. Like the original Skyrme model, it consists of a quadratic and a quartic term in derivatives, but in this case the presence of a zeroth derivative term (a potential) is followed by Derrick's argument if we want to ensure topological stable solutions. The target space manifold in the case of the bS model is $\mathbb{S}^2$ instead of $SU(2)$ and consequently, static configurations are maps from the one-point compactification of $\mathbb{R}^2$ to $\mathbb{S}^2$, and therefore they are classified, as in the original Skyrme proposal by a winding number. This reduced model, thought as a planar analogue of the Skyrme model, with similar topological properties can contribute to the understanding of the latter, but also has its own applications in different areas in theoretical physics, for example in  condensed matter \cite{Sondhi} and \cite{Schwindt}, or cosmology \cite{Sawado1,Sawado2,Sawado3}. 

One interesting, less explored, facet of these models is the issue of supersymmetry. Some years ago there were attempts to construct a supersymmetric version of a Skyrme-like model: $\mathbb{C}P^1$ Skyrme \cite{Nepo} and Faddeev-Skyrme \cite{Frey}. In both cases the SUSY version proposed contained terms with fourth time derivatives. More recently in \cite{queiruga1} and \cite{queiruga2} the first $N=1$ and $N=2$ SUSY extensions of baby Skyrme models were constructed, in \cite{Nitta1} the SUSY extension of the Faddeev-Skyrme model, in \cite{queiruga3} the $\mathbb{C}P^1$ Skyrme, and in \cite{Nitta2} the SUSY extension of the Skyrme model. Supersymmetric extensions of general theories containing higher derivative terms were constructed in \cite{Petrov,Queidef}. It is also interesting to note that the prototypical supersymmetric form of the $N=2$ bS model has four dimensional analogues, e.g. SUSY ghost condensates \cite{Khoury}, cosmic strings \cite{Trodden}, higher derivative supergravity \cite{Koehn} and SUSY galileons \cite{Ovrut}. In these cases, the $N=1$ SUSY action consists of a quadratic term in superfields (the trivial K\"ahler potential in four dimensions) plus a quartic term in superderivatives, which corresponds to the action of the bS model after dimensional reduction (note that in the reduction the supersymmetry is enlarged from $N=1$ to $N=2$).

Regarding the bS model, SUSY provides a natural way of introducing fermions. Once we have the SUSY version of a given model (the one which reduces to the model on its bosonic sector) we have at the same time the fermionic sector by the same price. But the SUSY extension of noncanonical kinetic terms (i.e. with more than two derivatives) is a nontrivial issue, and, even with the superfield formulation at hand, the study of the fermionic sector can be extremely complicated (see for example \cite{Trodden} for the explicit calculation of the fermionic sector of a fourth superderivative term). Maybe because of this reason the study of the fermions in these noncanonical models (in particular de bS), remains not too much developed. It is the main purpose of this work to give the first steps in the analysis of the fermions in the bS model.

This work is organized as follows. In Sec. II we present the full bS model and its BPS restriction [the so-called BPS baby Skyrme model (BbS)]. In Secs. III and IV we discuss the SUSY version of the model (obtained in \cite{queiruga2}) and some considerations relating SUSY and BPS equations are treated. In Sec. V we calculate the fermionic sector and in Sec. VI we exploit the supersymmetry of the model to obtain the fermionic zero modes in the background of the BPS solitons. In Sec. VII we show explicit examples and in Sec. VIII we discuss the consequences of a D-term deformation. Finally Sec. IX is devoted to our summary. We add also three appendices with notation and conventions, a brief discussion with other SUSY extensions and some considerations about the existence of compactons in models with D-terms.

%%%%%%%%%%%%%%%%%%%%%%%%%%%%%%%%%%%%%%%%%%%
%%%%%%%%%%%%%%%%%%%%%%%%%%%%%%%%%%%%%%%%%%%
%%%%%%%%%%%%%%%%%%%%%%%%%%%%%%%%%%%%%%%%%%%
%%%%%%%%%%%%%%%%%%%%%%%%%%%%%%%%%%%%%%%%%%%
%%%%%%%%%%%%%%%%%%%%%%%%%%%%%%%%%%%%%%%%%%%
%%%%%%%%%%%%%%%%%%%%%%%%%%%%%%%%%%%%%%%%%%%
%%%%%%%%%%%%%%%%%%%%%%%%%%%%%%%%%%%%%%%%%%%
%%%%%%%%%%%%%%%%%%%%%%%%%%%%%%%%%%%%%%%%%%%
%%%%%%%%%%%%%%%%%%%%%%%%%%%%%%%%%%%%%%%%%%%

\section{The model}

The Lagrangian of the baby Skyrme (bS) model can be written as the sum of three terms
\be
\mathcal{L}_{bS}=\mathcal{L}_2+\mathcal{L}_4+\mathcal{L}_0
\ee
where the subindex indicates the number of derivatives. The quadratic term corresponds to the usual $O(3)$ sigma model in three dimensions
\be
\mathcal{L}_2=\frac{\lambda_2}{4}\pa_\mu \vec{\phi}\pa^\mu\vec{\phi}.
\ee

The field $\vec{\phi}$ is a three component vector in $\mathbb{S}^2$. The potentials we are interested in depend only on the third component of the field, therefore $\mathcal{L}_0=-\lambda_0\mathcal{V}(\phi^3)$. Finally, the fourth derivative term can be expressed as
\be
\mathcal{L}_4=-\frac{\lambda_4}{8}\left(\pa_\mu\vec{\phi}\times\pa_\nu\vec{\phi}\right)^2.
\ee

The model possesses a Bogomol'nyi bound given by the topological charge, and also solutions saturating the bound when the $O(3)$ term is absent \cite{adam}-\cite{Stepien}. The later situation defines the so-called BPS baby Skyrme model (BbS) 
\be
\mathcal{L}_{BbS}=-\frac{\lambda_4}{8}\left(\pa_\mu\vec{\phi}\times\pa_\nu\vec{\phi}\right)^2-\lambda_0\mathcal{V}(\phi^3).
\ee

We can rewrite the Lagrangian in terms of the complex field $u$ after solving the constraint on $\vec{\phi}$ with the stereographic map
\be
\vec{\phi}=\frac{1}{1+\vert u\vert^2}\left(u+\bar{u},-i\left(u-\bar{u}\right),1-\vert u \vert^2\right).
\ee 

Taking $\lambda_4=\lambda_0=1$, the Lagrangian takes the following form in the new variables
\be
\mathcal{L}_{BbS}=-\frac{1}{\left(1+u\bar{u}\right)^4}\left(\left(\pa_\mu u\pa^\mu\bar{u}\right)^2-\left(\pa_\mu u\right)^2\left(\pa_\nu \bar{u}\right)^2\right)-\mathcal{V}\left( u\bar{u}\right).
\ee

These complex variables provide a more natural relation with the chiral superfields in the $N=2$ formulation of the model. One interesting feature of the model is that it has either nontrivial solutions with infinite and finite support (compactons) \cite{adam}, which we will analyze in the context of the supersymmetric model. In the symmetric ansatz $u=e^{i n \varphi}f(r)$ we have to impose the following boundary conditions
\be
f(r=0)=\infty,\quad f(r=R)=0\quad\text{and}\quad f'(r=R)=0\label{bound}
\ee  
where R stands for the size of the compacton ($R=\infty$ for noncompact solutions). We will see in the next section that these conditions can determine the breaking/preservation of a fraction of supersymmetry inside/outside of the defect solution.

%%%%%%%%%%%%%%%%%%%%%%%%%%%%%%%%%%%%%%%%%%%
%%%%%%%%%%%%%%%%%%%%%%%%%%%%%%%%%%%%%%%%%%%
%%%%%%%%%%%%%%%%%%%%%%%%%%%%%%%%%%%%%%%%%%%
%%%%%%%%%%%%%%%%%%%%%%%%%%%%%%%%%%%%%%%%%%%
%%%%%%%%%%%%%%%%%%%%%%%%%%%%%%%%%%%%%%%%%%%
%%%%%%%%%%%%%%%%%%%%%%%%%%%%%%%%%%%%%%%%%%%
%%%%%%%%%%%%%%%%%%%%%%%%%%%%%%%%%%%%%%%%%%%
%%%%%%%%%%%%%%%%%%%%%%%%%%%%%%%%%%%%%%%%%%%
%%%%%%%%%%%%%%%%%%%%%%%%%%%%%%%%%%%%%%%%%%%
\section{Supersymmetric baby Skyrme model}

The first $N=1$ supersymmetric extension of the bS was proposed in \cite{queiruga1}. If we demand one supersymmetry the quartic term can be supersymmetrized independently, and this implies in particular that both bS and BbS models possess at least an $N=1$ extension. The situation becomes more interesting if we demand two supersymmetries. First of all, in three dimensions and with two supersymmetries, there are chiral and antichiral complex superfields (in this dimension, the $N=1$ superfield formulation does not allow for such an object), so a natural guess for our superfield action is given by (see Appendix A for conventions)
\be
\mathcal{L}_{kin}=\int d^4 \theta K(\Phi,\Phi^\dagger)+\int d^4 \theta H(\Phi,\Phi^\dagger)D^\alpha\Phi D_\alpha \Phi \bar{D}^{\dot{\beta}}\Phi^\dagger \bar{D}_{\dot{\beta}}\Phi^\dagger\label{susyact}
\ee

and 
\be
\mathcal{L}_{pot}=\int d^2\theta \,W(\Phi)+\text{H.c.}\label{susypot}
\ee

The first term in (\ref{susyact}) is a K\"ahler potential and it generates the usual nonlinear $\sigma$-model term. The second term generates a fourth derivative term and (\ref{susypot}) stands for the usual prepotential term. We will see later that the presence of $\mathcal{L}_{pot}$ leads to an action with exotic dynamics (which does not correspond to the BbS model). It could seem that, neglecting the prepotential term, we have no chance to generate a potential in the bosonic sector, and thus breaking the stability of the model. We will see that this is in fact not true. After integration in the Grassmann coordinates in (\ref{susyact}) and switching off fermions we get \cite{queiruga2}
\be
\mathcal{L}_{kin}^{bos}=g(u,\bar{u})\left( \pa^\mu u\pa_\mu\bar{u}+F\bar{F}\right)+h(u,\bar{u})\left((\pa_\mu u)^2 (\pa_\nu \bar{u})^2+2 F\bar{F} \pa^\mu\bar{u}\pa_\mu u +\left(F\bar{F}\right)^2\right)\label{bosact}
\ee
where $g(u,\bar{u})$ in the K\"ahler metric coming from the K\"ahler potential:
\be
g(u,\bar{u})=\frac{\pa^2}{\pa\Phi\pa\Phi^\dagger}K(\Phi,\Phi^\dagger)\vert_{\theta=\bar{\theta}=0}
\ee

and
\be
h(u,\bar{u})=H(\Phi,\Phi^\dagger)\vert_{\theta=\bar{\theta}=0}.
\ee

It remains to eliminate the auxiliary field $F$. The trivial solution $F=0$ leads to a $\sigma$-model term plus a quartic term (which does not correspond to the bS or BbS models). Fortunately there is one extra solution
\be
F=e^{i\eta}\sqrt{-\pa_\mu u\pa^\mu\bar{u}-\frac{g(u,\bar{u})}{2h(u,\bar{u})}}\label{aux}
\ee
where $\eta$ is an arbitrary phase [note that the Lagrangian (\ref{bosact}) is invariant under the replacement $F\rightarrow e^{i\eta}F$]. Plugging this solution into (\ref{bosact}) we obtain
\be
\mathcal{L}_{kin}^{bos}=h(u,\bar{u})\left((\pa_\mu u)^2 (\pa_\nu \bar{u})^2-(\pa^\mu\bar{u}\pa_\mu u )^2\right)-\frac{g(u,\bar{u})^2}{4 h(u,\bar{u})}
\ee
after the choice $h(u,\bar{u})=1/(1+u\bar{u})^4$ we arrive at the Lagrangian of the BbS model with potential
\be
\mathcal{V}(u,\bar{u})=\frac{1}{4}g(u,\bar{u})^2(1+u\bar{u})^4.
\ee

Three observations are in order. First, the potential of the model is completely determined by the K\"ahler metric coming from the $\sigma$-model part, so, no need for introducing a chiral prepotential. Second, the non-trivial solution for $F$ (\ref{aux}) eliminates the quadratic term from the action, and therefore we are free to choose the K\"ahler potential $K$, without changing the kinetic terms.  Third, it seems that the model including the quadratic term does not allow for two supersymmetries since the $\sigma$-model term was "eaten" by the auxiliary field $F$, however, we will see later that other $N=2$ supersymmetric extensions of the bS exist.

%%%%%%%%%%%%%%%%%%%%%%%%%%%%%%%%%%%%%%%%%%%
%%%%%%%%%%%%%%%%%%%%%%%%%%%%%%%%%%%%%%%%%%%
%%%%%%%%%%%%%%%%%%%%%%%%%%%%%%%%%%%%%%%%%%%
%%%%%%%%%%%%%%%%%%%%%%%%%%%%%%%%%%%%%%%%%%%
%%%%%%%%%%%%%%%%%%%%%%%%%%%%%%%%%%%%%%%%%%%
%%%%%%%%%%%%%%%%%%%%%%%%%%%%%%%%%%%%%%%%%%%
%%%%%%%%%%%%%%%%%%%%%%%%%%%%%%%%%%%%%%%%%%%
%%%%%%%%%%%%%%%%%%%%%%%%%%%%%%%%%%%%%%%%%%%
%%%%%%%%%%%%%%%%%%%%%%%%%%%%%%%%%%%%%%%%%%%
\section{BPS equation and supersymmetry}

The energy functional of the model can be written as follows:
\be
E=\int d^2x\left( \frac{\left(\pa_i u\pa^i \bar{u}\right)^2-\left(\pa_i u\right)^2\left(\pa_j \bar{u}\right)^2}{\left(1+u\bar{u}\right)^4} +\mathcal{V}(u,\bar{u})\right).
\ee

For our future purposes it is more convenient to write the energy functional in the following way
\be
E=\frac{1}{4}\int d^2 x\left( \frac{\left(\vert\pa u\vert^2-\vert\bar{\pa}u\vert^2\right)^2}{\left(1+u\bar{u}\right)^4}+4\mathcal{V}(u,\bar{u})\right)\label{ener}
\ee
where $\pa$ is the holomorphic derivative defined by $\pa=\pa_1+i\pa_2$. We can split (\ref{ener}) into a square and other term
\be
E=\frac{1}{4}\int d^2 x \left\{ \left( \frac{\left(\vert\pa u\vert^2-\vert\bar{\pa}u\vert^2\right)}{\left(1+u\bar{u}\right)^2}\pm2\sqrt{\mathcal{V}(u,\bar{u})}\right)^2\mp4\sqrt{V(u,\bar{u})}\frac{\left(\vert\pa u\vert^2-\vert\bar{\pa}u\vert^2\right)}{\left(1+u\bar{u}\right)^2} \right\},\label{enersplit}
\ee

therefore
\be
E\geq 	\mp 4\pi	\int d^2 x \sqrt{V(u,\bar{u})}Q, \quad Q=\frac{1}{4\pi}\frac{\left(\vert\pa u\vert^2-\vert\bar{\pa}u\vert^2\right)}{\left(1+u\bar{u}\right)^2} 
\ee
(Q is the topological charge density). The expression for the BPS equation can be read from (\ref{enersplit}) and after some manipulations we get
\be
\pa u=e^{i\eta}\sqrt{\sqrt{V(u,\bar{u})}\left(1+u\bar{u}\right)^2+\frac{1}{2}\left(\vert\pa u\vert^2+\vert\bar{\pa}u\vert^2\right)}\label{BPS}
\ee
where $\eta$ is an arbitrary phase [note that the Lagrangian (\ref{bosact}) is invariant under $U(1)$ global transformation on both $u$ and $F$ fields]. We can write (\ref{BPS}) in a more suggestive way taking into account the on-shell solution for the auxiliary field (\ref{aux}) in the supersymmetric version of the model:
\be
\pa u=e^{i\eta'}F.\label{bpssusy}
\ee

This expression can be obtained directly from SUSY transformations of the fermions
\be
\delta\psi^\alpha=\pa^{\alpha\dot{\alpha}}u\,\bar{\xi}_{\dot{\alpha}}+F\xi^\alpha
\ee
and its conjugate. We use $\pa_{\alpha\dot{\alpha}}=\sigma^1_{\alpha\dot{\alpha}}\pa_1+\sigma^3_{\alpha\dot{\alpha}}\pa_2$ (static regime).  Therefore preservation of (a fraction of) supersymmetry is equivalent to the condition
\be
 \delta\psi^\alpha\vert_{\text{on-shell}}=0\label{ferm}
\ee 
for some constant spinors $\xi,\bar{\xi}$. Since we have four supersymmetric generators ($N=2$), the fraction of supersymmetry which is preserved by BPS solutions will be given by $\dim \,\text{Ker}(\delta)$. The condition (\ref{ferm}) can be expanded as follows
\be
+\frac{1}{2}\left(\sigma_1^{\alpha\dot{\beta}}+i\sigma_3^{\alpha\dot{\beta}}\right)\pa u\, \bar{\xi}_{\dot{\beta}}-\frac{1}{2}\left(\sigma_1^{\alpha\dot{\beta}}-i\sigma_3^{\alpha\dot{\beta}}\right)\bar{\pa}u\,\bar{\xi}_{\dot{\beta}}+F\xi^\alpha=0.\label{bppp}
\ee

In the light of (\ref{bppp}) we deduce immediately the following Killing conditions for the supersymmetric parameters
\be
\left(\sigma_1^{\alpha\dot{\beta}}-i\sigma_3^{\alpha\dot{\beta}}\right)\,\bar{\xi}_{\dot{\beta}}=0\quad \text{and} \quad \frac{1}{2}\left(\sigma_1^{\alpha\dot{\beta}}+i\sigma_3^{\alpha\dot{\beta}}\right)\, \bar{\xi}_{\dot{\beta}}=-e^{i\eta'}\xi^{\alpha}\label{cons}
\ee 

The constraints (\ref{cons}) reduce the dimensions of the static superspace from $(2\vert4)$ to $(2\vert 1)$, or equivalently $\dim \,\text{Ker}(\delta)\vert_{\text{on-shell}}=1$ and therefore, solutions satisfying (\ref{bpssusy}) are 1/4 supersymmetric. This can be confirmed by calculating the eigenvalues of the operator $\delta$. We obtain the following results
\bea
\lambda_1^{\pm}&=&\text{Re}\,F\pm\sqrt{\bar{\pa}u\pa\bar{u}-\text{Im}^2\,F}\label{d1}\\
\lambda_2^{\pm}&=&\text{Re}\,F\pm\sqrt{\pa u\bar{\pa}\bar{u}-\text{Im}^2\,F}.\label{d2}
\eea

 If we substitute (\ref{bpssusy}) in (\ref{d1}) and (\ref{d2}) the only vanishing eigenvalue is $\lambda_2^-$ which implies $\text{Rank}\, \delta\vert_{\text{on-shell}}=3\Leftrightarrow \dim \,\text{Ker}(\delta)\vert_{\text{on-shell}}=1$. This result holds for generic solutions of the BPS equation, but we will see that, in the case of compactons, supersymmetry is restored out of the defect. 

%%%%%%%%%%%%%%%%%%%%%%%%%%%%%%%%%%%%%%%%%%%
%%%%%%%%%%%%%%%%%%%%%%%%%%%%%%%%%%%%%%%%%%%
%%%%%%%%%%%%%%%%%%%%%%%%%%%%%%%%%%%%%%%%%%%
%%%%%%%%%%%%%%%%%%%%%%%%%%%%%%%%%%%%%%%%%%%
%%%%%%%%%%%%%%%%%%%%%%%%%%%%%%%%%%%%%%%%%%%
%%%%%%%%%%%%%%%%%%%%%%%%%%%%%%%%%%%%%%%%%%%
%%%%%%%%%%%%%%%%%%%%%%%%%%%%%%%%%%%%%%%%%%%
%%%%%%%%%%%%%%%%%%%%%%%%%%%%%%%%%%%%%%%%%%%
%%%%%%%%%%%%%%%%%%%%%%%%%%%%%%%%%%%%%%%%%%%
\section{fermionic sector}

Despite the apparent simplicity of the bosonic sector of the model, the analysis of the fermionic sector is highly nontrivial, even if we consider the Lagrangian up to quadratic order in the fermionic field. The first contribution is very well known and corresponds to the usual $N=2$ supersymmetric nonlinear sigma model. After integration we get

\be
\mathcal{L}_2=\int d^2\theta d^2\bar{\theta}K(\Phi,\Phi^\dagger)=g(u,\bar{u})\left(\pa^\mu u\pa_\mu\bar{u}-\frac{i}{2}\psi \sigma^\mu \mathcal{D}^\mu \bar{\psi}+\frac{i}{2}\mathcal{D}^\mu\psi\sigma_\mu \bar{\psi}+F\bar{F}\right)+\mathcal{O}(\psi\bar{\psi})^2
\ee
where $g(u,\bar{u})$ is the K\"ahler metric and $\mathcal{D}_\mu$ is the covariant derivative defined by
\be
D_\mu=\pa_\mu+\Gamma^u_{uu}\pa_\mu, \quad \Gamma^u_{uu}=g^{-1}(u,\bar{u})\pa_u g(u,\bar{u}).
\ee

We can split the quartic term according to the number of derivatives on the function $H(\Phi,\Phi^\dagger)$. The term involving zero derivatives takes the following form
\bea
\mathcal{L}_4^0&=&h(u,\bar{u})\left( (\pa_\mu u)^2(\pa_\nu \bar{u})^2+2 F\bar{F}\pa^\mu u\pa_\nu\bar{u}+(F\bar{F})^2 +\frac{i}{2}\psi\sigma^\mu\bar{\psi}\left(\bar{u}_{,\mu}\square u-u_{,\mu}\square\bar{u}\right)  \right.\nonumber\\
&&\left.-\frac{i}{2}\psi\sigma^\mu\bar{\psi}^{,\nu}u_{,\mu}\bar{u}_{,\nu}-i\psi^{,\mu}\sigma^\nu\bar{\psi}u_{,\mu}\bar{u}_{,\nu}-\frac{i}{2}\psi\sigma^\mu\bar{\sigma}^\rho\sigma^\nu\bar{\psi}_{,\nu}u_{,\mu}\bar{u}_{,\rho}+\frac{i}{2}\psi_{,\nu}\sigma^\nu\bar{\sigma}^\mu\sigma^\rho u_{,\mu}\bar{u}_{,\rho} \right.\nonumber\\
&&\left.+F\square u\bar{\psi}\bar{\psi}+\frac{1}{2}F u_{,\mu} \pa^\mu(\bar{\psi}\bar{\psi})+\bar{F}\square \bar{u}\psi\psi+\frac{1}{2}\bar{F}\bar{u}_{,\mu}\pa^\mu (\psi\psi) \right.\nonumber\\
&&\left.+\frac{1}{2}F u_{,\mu}\left(\psi \bar{\sigma}^\mu\sigma^\nu\bar{\psi}_{,\nu}-\bar{\psi}_{,\nu}\bar{\sigma}^\mu\sigma^\nu\bar{\psi}\right) +\frac{1}{2}\bar{F}\bar{u}_{,\mu}\left(\psi_{,\nu}\sigma^\nu\bar{\sigma}^\mu \psi-\psi\sigma^\nu\bar{\sigma}^\mu \psi_{,\nu}\right)\right.\nonumber\\
&&\left.+\frac{3i}{2}F\bar{F}\left(\psi_{,\mu}\sigma^\mu\bar{\psi}-\psi\sigma^\mu\bar{\psi}_{,\mu}\right)+\frac{i}{2}\psi \sigma^\mu \bar{\psi}\left(F\bar{F}_{,\mu}-\bar{F}F_{,\mu}\right)+\mathcal{O}(\psi)^3 \right).\label{full1}
\eea

%The term with one derivative is:
%\bea
%\mathcal{L}_4^{1}&=&\pa_u h(u,\bar{u})\left(\frac{1}{2}F\bar{F^2}\psi^2-\frac{1}{2}F\psi^2(\bar{u}_{,\nu})^2+i\sigma^\mu\bar{\psi}\pa_\mu\psi -i\sqrt{2}(\bar{u}_{,\mu})^2\bar{\sigma}^\nu \psi^2 u_{,\nu}  \right.\nonumber\\
%&&\left.-2i F\bar{F} \bar{u}_{,\mu}\psi\sigma^\mu\psi-2 F\bar{u}_{,\mu}\psi\bar{\psi}\bar{\sigma}^\mu\sigma_\nu u_{,\nu} +2i \psi\bar{\psi}\sigma^\mu\sigma^\nu\sigma^\rho u_{,\mu} \bar{u}_{,\nu} u_{,\rho} \right.\nonumber\\
%&&\left.+2 F\bar{F}\psi \sigma^\mu \bar{\psi} u_{,\mu}-F\bar{\psi}^2 \sigma^\mu\sigma^\nu u_{,\mu} u_{,\mu}+\bar{F}\psi^2 \sigma^\mu\sigma^\nu \bar{u}_{,\mu}u_{,\nu}+\mathcal{O}(\psi)^3\right).
%\eea

%And finally the term with two derivatives can be written as:

%\bea
%\mathcal{L}_4^{2}&=&\pa_{u,\bar{u}}^2h(u,\bar{u})\left(\pa^\mu u\pa_\mu\bar{u}-\frac{i}{2}\psi \sigma^\mu \tilde{\mathcal{D}}^\mu \bar{\psi}+\frac{i}{2}\tilde{\mathcal{D}}^\mu\psi\sigma_\mu \bar{\psi}+F\bar{F}\right)i\psi\sigma^\mu\bar{\psi}(F \bar{u}_{,\mu}-\bar{F}u_{,\nu})\nonumber\\
%&+&\pa_{u,\bar{u}}^2h(u,\bar{u})\left(-i\sigma^\mu\bar{u}_{,\mu}\bar{\psi}\psi-2F\bar{\psi}\bar{\psi}\right)+\pa_u^2 h(u,\bar{u})2\bar{\psi}\sigma^\mu\psi u_{,\mu}-\pa_{\bar{u}}^2 h(u,\bar{u})2\psi\sigma^\mu\bar{\psi} \bar{u}_{,\mu}
%\eea
%where $\tilde{\mathcal{D}}_\mu$ is the covariant derivative defined with respect to $h(u,\bar{u})$. 

The remaining terms involving one and two derivatives with respect to the function $h(u,\bar{u})$ are rather cumbersome and not very enlightening for our purposes. If we denote $\mathcal{L}_4^1$ and $\mathcal{L}_4^2$  these two terms then the full Lagrangian is
\be
\mathcal{L}=\mathcal{L}_2+\mathcal{L}_4^0+\mathcal{L}_4^1+\mathcal{L}_4^{1\dagger}+\mathcal{L}_4^2.\label{full}
\ee

 If we switch off fermions in (\ref{full}) and take the nonzero solutions for $F$ the resulting Lagrangian corresponds to the BbS model as we discussed above. It seems also that in the full model the auxiliary field becomes dynamical and satisfies a first order differential equation as we can see from the last term of (\ref{full1}). Moreover, we cannot even ensure that the solution $F=0$ satisfies the equation of motion due to the linear terms in $F$ which show up in the fermionic sector. It seems therefore that the study of fermions in the model remains intractable by brute force. Fortunately we can exploit the supersymmetry of the model to obtain some information about fermion zero modes from the study of bosonic solutions as we show in the next section.

%%%%%%%%%%%%%%%%%%%%%%%%%%%%%%%%%%%%%%%%%%%
%%%%%%%%%%%%%%%%%%%%%%%%%%%%%%%%%%%%%%%%%%%
%%%%%%%%%%%%%%%%%%%%%%%%%%%%%%%%%%%%%%%%%%%
%%%%%%%%%%%%%%%%%%%%%%%%%%%%%%%%%%%%%%%%%%%
%%%%%%%%%%%%%%%%%%%%%%%%%%%%%%%%%%%%%%%%%%%
%%%%%%%%%%%%%%%%%%%%%%%%%%%%%%%%%%%%%%%%%%%
%%%%%%%%%%%%%%%%%%%%%%%%%%%%%%%%%%%%%%%%%%%
%%%%%%%%%%%%%%%%%%%%%%%%%%%%%%%%%%%%%%%%%%%
%%%%%%%%%%%%%%%%%%%%%%%%%%%%%%%%%%%%%%%%%%%

\section{Zero modes}

Let us define the following spinors in terms of the original Grassmann parameters of the supersymmetric transformations:
\bea
\eta^\alpha&=&-\frac{1}{2}\left(\sigma_1^{\alpha\dot{\beta}}-i\sigma_3^{\alpha\dot{\beta}}\right)\,\bar{\xi}_{\dot{\beta}}\\
\rho^\alpha&=&+\frac{1}{2}\left(\sigma_1^{\alpha\dot{\beta}}+i\sigma_3^{\alpha\dot{\beta}}\right)\, \bar{\xi}_{\dot{\beta}}+e^{i\eta'}\xi^{\alpha}.
\eea

In terms of the new variables the conditions (\ref{cons}) can be written simply as $\rho^\alpha=\eta^\alpha=0$ and the supersymmetric transformations of the fermions as

\be
\delta\psi^\alpha=\rho^\alpha \pa u+ \eta^\alpha\bar{\pa}u+\left(e^{i\eta}\pa u+F\right)\xi^\alpha\label{fermt}
\ee

The on-shell BPS transformation verifies
\be
\delta\psi^\alpha\vert_{\text{BPS}}=\rho^\alpha \pa u+ \eta^\alpha\bar{\pa}u\label{trans}
\ee

It is important to note that the spinors $\rho^\alpha$ and $\eta^\alpha$ have two and one Grassmannian degrees of freedom, respectively, once $\xi^\alpha$ is absent in (\ref{trans}). As a consequence, the fermionic zero modes in the background on the BPS solution will have three degrees of freedom (corresponding to the three broken supercharges)
\bea
\eta^\alpha&=&- \frac{1}{2}\chi\left(\begin{matrix}
    1   \\
     i
\end{matrix}\right),\quad \chi=i\bar{\xi}_{\dot{1}}-\bar{\xi}_{\dot{2}}\\
\rho^\alpha&=&-\frac{1}{2}\left(\begin{matrix}
    \zeta_1   \\
     \zeta_2
\end{matrix}\right),\quad \zeta_1=-i\bar{\xi}_{\dot{1}}-\bar{\xi}_{\dot{2}}-2e^{i\eta}\xi^1,\quad  \zeta_2=\bar{\xi}_{\dot{1}}+i\bar{\xi}_{\dot{2}}-2e^{i\eta}\xi^2.
\eea

 See for example \cite{Abraham,Vainshtein, Ito} for the explicit calculation of fermionic zero modes in different models. We can split the fermionic zero modes into three elements
\be
\left\{ \frac{1}{2}\chi\left(\begin{matrix}
    1   \\
     i
\end{matrix}\right)\bar{\pa}u,\frac{1}{2}\left(\begin{matrix}
    \zeta_1   \\
    0
\end{matrix}\right)\pa u,\frac{1}{2}\left(\begin{matrix}
    0   \\
     \zeta_2
\end{matrix}\right) \pa u\right\}.
\ee

Now if we assume the symmetric ansatz $u=f(r)e^{in\varphi}$ for the BPS solutions we obtain:
\be
\delta\psi^\alpha\vert_{\text{BPS}}=\rho^\alpha\frac{e^{i (n+1) \varphi } \left(r f'(r)-n f(r)\right)}{r}+ \eta^\alpha \frac{e^{i (n-1) \theta } \left(r f'(r)+n f(r)\right)}{r}\label{trans1}
\ee

and for the zero modes

%\be
%\left\{ \frac{1}{2}\chi\left(\begin{matrix}
    %1   \\
    % i
%\end{matrix}\right) \frac{e^{i (n-1) \theta } \left(r f'(r)+n f(r)\right)}{r},\frac{1}{2}  \zeta_1\left(\begin{matrix}
 % 1   \\
    %0
%\end{matrix}\right)\frac{e^{i (n+1) \varphi } \left(r f'(r)-n f(r)\right)}{r},\\\frac{1}{2}  \zeta_2\left(\begin{matrix}
 %   0   \\
    %1
%\end{matrix}\right)\frac{e^{i (n+1) \varphi } \left(r f'(r)-n f(r)\right)}{r}\right\}\label{zero}
%\ee
\be
\left(\begin{matrix}
\frac{1}{2}\chi\left(\begin{matrix}
    1   \\
     i
\end{matrix}\right) \frac{e^{i (n-1) \theta } \left(r f'(r)+n f(r)\right)}{r}\\ \frac{1}{2}\zeta_1\left(\begin{matrix}
   1   \\
    0
\end{matrix}\right)\frac{e^{i (n+1) \varphi } \left(r f'(r)-n f(r)\right)}{r}\\\frac{1}{2}\zeta_2\left(\begin{matrix}
    0   \\
    1
\end{matrix}\right)\frac{e^{i (n+1) \varphi } \left(r f'(r)-n f(r)\right)}{r}
\end{matrix}\right)\label{zero}
\ee

The expression (\ref{zero}) has the following interesting consequence. Let us take a BPS solution satisfying the boundary conditions (\ref{bound}) with $R=\infty$, i.e. an infinitely extended BPS baby Skyrmion. Under this assumption $\delta\psi^\alpha\vert_{\text{BPS}}\neq 0 $ unless $\rho^\alpha=\eta^\alpha=0$ therefore, it preserves only 1/4 supersymmetry in $\mathbb{R}^2$, as pointed out  above, and the fermionic zero modes (\ref{zero}) exist throughout space. But if we take a compacton, since $f'(r)=0$ and $f(r)=0,r\geq R$, the transformation of the fermions also vanishes in the region $r\geq R$ and  fermionic zero modes and partial breaking of SUSY are confined to the defect core. In the following section we study explicit examples.

%%%%%%%%%%%%%%%%%%%%%%%%%%%%%%%%%%%%%%%
%%%%%%%%%%%%%%%%%%%%%%%%%%%%%%%%%%%%%%%
%%%%%%%%%%%%%%%%%%%%%%%%%%%%%%%%%%%%%%%
%%%%%%%%%%%%%%%%%%%%%%%%%%%%%%%%%%%%%%%
%%%%%%%%%%%%%%%%%%%%%%%%%%%%%%%%%%%%%%%
%%%%%%%%%%%%%%%%%%%%%%%%%%%%%%%%%%%%%%%
%%%%%%%%%%%%%%%%%%%%%%%%%%%%%%%%%%%%%%%

\section{Compactons and compact fermions}

Following \cite{adam} we introduce the following field variable $g$ with respect to the symmetric ansatz
\be
1-g=\frac{1}{1+f^2}.
\ee

The boundary conditions (\ref{bound}) in the new variable read
\be
g(r=0)=1,\quad g(r=R)=0\quad \text{and}\quad g'(r=R)=0.
\ee

If we consider the potential term
\be
\mathcal{V}(u,\bar{u})=\left(\frac{u\bar{u}}{1+u\bar{u}}\right)^s\label{pot}
\ee
the corresponding K\"ahler potential for the supersymmetric version of the model takes the form
\be
K(\Phi,\Phi^\dagger)=8\frac{( \Phi\Phi^\dagger )^{\frac{s+2}{2}}}{(s+2)^2} \,_2 F_1\left(\frac{s+2}{2},\frac{s+2}{2},\frac{s+4}{2},-\Phi^\dagger \Phi   \right).\label{kpote3}
\ee

Note that to obtain the expression (\ref{kpote3}) we only need to integrate the K\"ahler metric (given by the form of the potential) twice with respect to the holomorphic/antiholomorphic field variables and replace them with the chiral/antichiral superfields. The compact BPS solutions which correspond to the potential (\ref{pot}) are given by the following expression
\be
g(r)= \begin{cases} 
      (1-\frac{r^2}{R^2})^{\frac{2}{2-s}}, & 0\leq r\leq R \\
      0 ,& r\geq R 
   \end{cases}
\ee

where $s\in(1,2)$ and the compacton boundary is given by
\be
R^2=\frac{4 n}{2-s}.
\ee

Returning to the variable $f$ we obtain for $n=1$
\be
f(r)=\begin{cases} 
\frac{\left(1-\frac{r^2}{R^2}\right)^{\frac{1}{2-s}}}{\sqrt{1-\left(1-\frac{r^2}{R^2}\right)^{-\frac{2}{s-2}}}},& 0\leq r\leq R \\
   0, & r\geq R 
   \end{cases}
\ee
and
\be
f'(r)=\begin{cases} 
-\frac{2 r \left(1-\frac{r^2}{R^2}\right)^{\frac{1}{2-s}}}{(s-2) \left(r^2-R^2\right)
   \left(1-\left(1-\frac{r^2}{R^2}\right)^{-\frac{2}{s-2}}\right)^{3/2}},& 0\leq r\leq R \\
   0 ,& r\geq R .
   \end{cases}
\ee

\begin{figure}[h]
    \centering
    \includegraphics[width=1.0\textwidth]{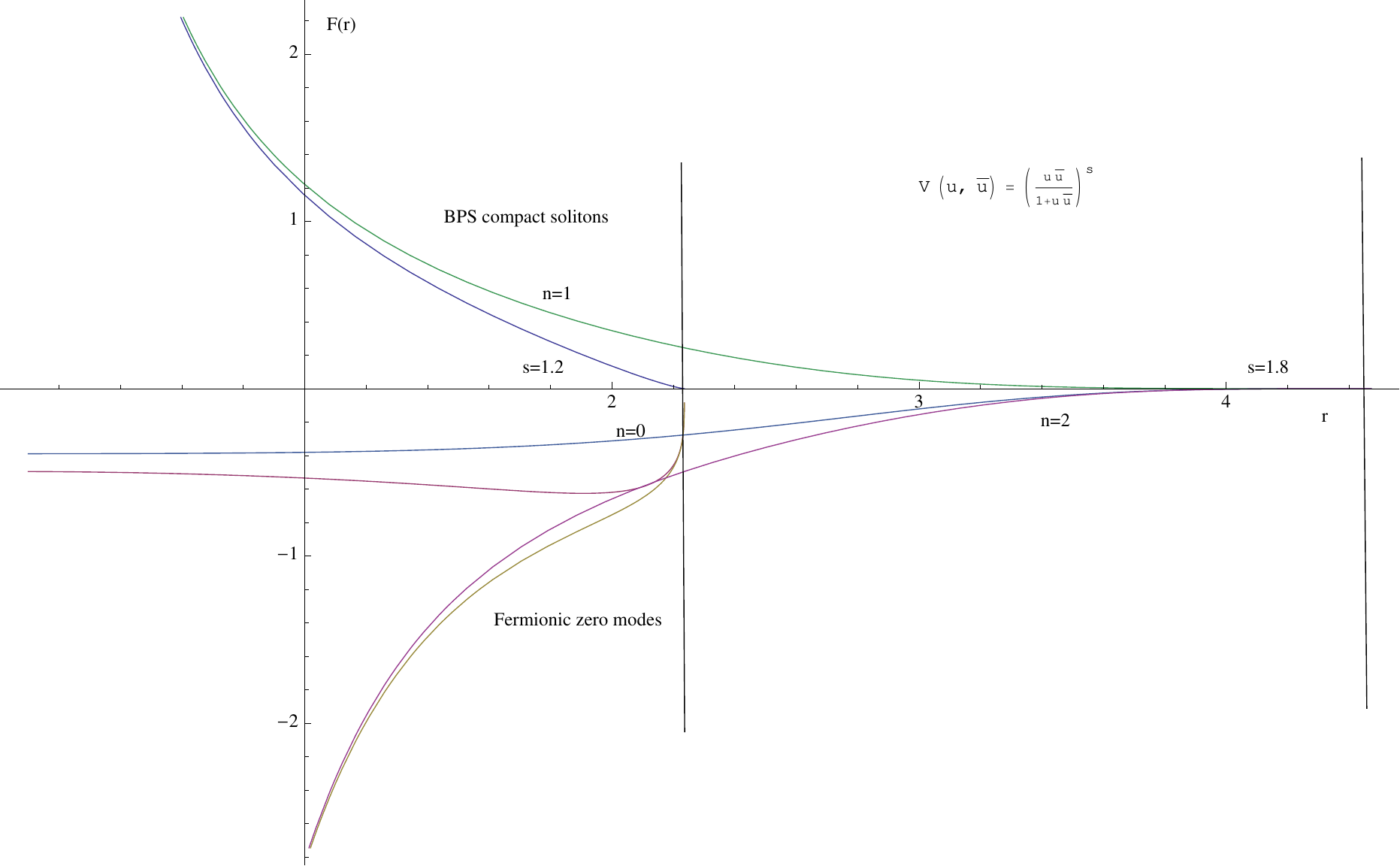}
    \caption{BPS compact solitons and fermionic zero modes}
    \label{fig1}
\end{figure}

In Fig. \ref{fig1} we show the profile functions for different values of the exponent $s$. The curves in the upper half plane show $f(r)$ (the BPS compact solitons), while in the lower half plane are depicted the profiles of the fermionic zero modes which correspond to $F(r)\equiv\left(f'(r)\pm\frac{1}{r} f(r)\right)$ as can be seen in (\ref{zero}). As $s$ approaches to the critical value $s=2$ the support of the compacton grows. Outside of this support fermionic zero modes do not exist and full supersymmetry is recovered. For the value $s=2$ the BPS solitons are exponentially localized with infinite support, and therefore, fermionic zero modes are not confined to the defect core, but they extend throughout $\mathbb{R}^2$ space. This analysis can be straightforwardly generalized to a bigger family of potentials: nonsymmetric, vortex like... \cite{adam}, but as long as they allow for the existence of BPS solutions, the relation between the latter and fermionic zero modes remains unchanged (\ref{zero}).

%%%%%%%%%%%%%%%%%%%%%%%%%%%%%%%%%%%%%%%%%%%
%%%%%%%%%%%%%%%%%%%%%%%%%%%%%%%%%%%%%%%%%%%
%%%%%%%%%%%%%%%%%%%%%%%%%%%%%%%%%%%%%%%%%%%
%%%%%%%%%%%%%%%%%%%%%%%%%%%%%%%%%%%%%%%%%%%
%%%%%%%%%%%%%%%%%%%%%%%%%%%%%%%%%%%%%%%%%%%
%%%%%%%%%%%%%%%%%%%%%%%%%%%%%%%%%%%%%%%%%%%
%%%%%%%%%%%%%%%%%%%%%%%%%%%%%%%%%%%%%%%%%%%
%%%%%%%%%%%%%%%%%%%%%%%%%%%%%%%%%%%%%%%%%%%
%%%%%%%%%%%%%%%%%%%%%%%%%%%%%%%%%%%%%%%%%%%

\section{Exotic models with D- terms}

In this section we add a D-term (prepotential) to our previous action. In terms of superfields the new Lagrangian can be written as follows
\be
\mathcal{L}=\int d^4 \theta K(\Phi,\Phi^\dagger)+\int d^4 \theta H(\Phi,\Phi^\dagger)D^\alpha\Phi D_\alpha \Phi \bar{D}^{\dot{\beta}}\Phi^\dagger \bar{D}_{\dot{\beta}}\Phi^\dagger+\left(\lambda\int d^2\theta W (\Phi)+\text{H.c.}	\right).\label{susyact1}
\ee

After integration we obtain in the bosonic sector 
\bea
\mathcal{L}&=&g(u,\bar{u})\left( \pa^\mu u\pa_\mu\bar{u}+F\bar{F}\right)+h(u,\bar{u})\left((\pa_\mu u)^2 (\pa_\nu \bar{u})^2+2 F\bar{F} \pa^\mu\bar{u}\pa_\mu u +\left(F\bar{F}\right)^2\right)+\nonumber\\
&+&\lambda F W'(u)+\lambda\bar{F}\bar{W}'(\bar{u}).\label{bosex}
\eea

Now the equations of motion are cubic and the solutions, given by Cardano formulae are cumbersome. The equations of motion for the auxiliary fields take the following form
\bea
g(u,\bar{u}) F+2h(u,\bar{u})\left( F \pa^\mu\bar{u}\pa_\mu u+F (F\bar{F})\right)+\lambda \bar{W}'(\bar{u})&=&0\label{a1}\\
g(u,\bar{u}) \bar{F}+2h(u,\bar{u})\left( \bar{F} \pa^\mu\bar{u}\pa_\mu u+\bar{F} (F\bar{F})\right)+\lambda W'(u)&=&0\label{a2}.
\eea

From (\ref{a1}) and (\ref{a2}) we arrive to the following relation between $F$ and $\bar{F}$
\be
\bar{F}^{(i)}=\frac{W'}{\bar{W}'}F^{(i)}.
\ee

After reinserting this relation into (\ref{a1}) and expanding for $\lambda\sim0$ we obtain three different branches
% We can treat the D-term as a perturbation over the original action and expand in the parameter $\lambda$ for $\lambda\sim0$. We obtain three different branches
\bea
F^{(1)}&=&-\lambda \frac{\bar{W}'}{g(u,\bar{u})+2 h(u,\bar{u})u_{,\mu}\bar{u}^{,\mu}}+\mathcal{O}(\lambda^2)\\
F^{(2)}&=&-\sqrt{\frac{\bar{W}'}{W'}}\sqrt{-\frac{g(u,\bar{u})}{2h(u,\bar{u})}-u_{,\mu}\bar{u}^{,\mu}}+\frac{\lambda}{2} \frac{\bar{W}'}{g(u,\bar{u})+2 h(u,\bar{u})u_{,\mu}\bar{u}^{,\mu}}+\mathcal{O}(\lambda^2)\label{Daux2}\\
F^{(3)}&=&\sqrt{\frac{\bar{W}'}{W'}}\sqrt{-\frac{g(u,\bar{u})}{2h(u,\bar{u})}-u_{,\mu}\bar{u}^{,\mu}}+\frac{\lambda}{2} \frac{\bar{W}'}{g(u,\bar{u})+2 h(u,\bar{u})u_{,\mu}\bar{u}^{,\mu}}+\mathcal{O}(\lambda^2)
\eea
where $u_{,\mu}\equiv \pa_\mu u$. After substituting these values in (\ref{bosex}) we arrive at the following Lagrangians
\bea
\mathcal{L}^{(1)}&=&g(u,\bar{u})u_{,\mu}\bar{u}^{,\mu}+h(u,\bar{u})(u_{,\mu}^2)(\bar{u}_{,\nu})^2+\mathcal{O}(\lambda^2)  \label{ex1}\\
\mathcal{L}^{(2)}&=&h(u,\bar{u})\left(( u_{,\mu})^2 (\bar{u}_{,\nu})^2-(u_{,\mu}\bar{u}^{,\mu})^2\right)-\frac{g(u,\bar{u})^2}{4 h(u,\bar{u})}\nonumber\\
&&+2\lambda \vert W'\vert \sqrt{-\frac{g(u,\bar{u})}{2h(u,\bar{u})}-u_{,\mu}\bar{u}^{,\mu}}+\mathcal{O}(\lambda^2)  \label{ex2}  \\
\mathcal{L}^{(3)}&=&  h(u,\bar{u})\left(( u_{,\mu})^2 (\bar{u}_{,\nu})^2-(u_{,\mu}\bar{u}^{,\mu} )^2\right)-\frac{g(u,\bar{u})^2}{4 h(u,\bar{u})}\nonumber\\
&&-2\lambda \vert W'\vert \sqrt{-\frac{g(u,\bar{u})}{2h(u,\bar{u})}-u_{,\mu}\bar{u}^{,\mu}}+\mathcal{O}(\lambda^2). \label{ex3}
\eea

The first Lagrangian (\ref{ex1}) gives the usual nonlinear sigma model term plus a fourth derivative term, while the other two, (\ref{ex2}) and (\ref{ex3}) give the BbS model term plus a correction in $\lambda$ (those are connected with the nontrivial solutions of F which generate the BbS without D term). Let us assume that these models possess also compacton solutions verifying the boundary conditions (\ref{bound}) (see Appendix C). If these solutions with compact support are BPS then all the discussion of Sec. VII holds, if not, from (\ref{fermt}) (in the symmetric ansatz) we get
\bea
\delta\psi^\alpha&=&\rho^\alpha\frac{e^{i (n+1) \varphi } \left(r f'(r)-n f(r)\right)}{r}+ \eta^\alpha \frac{e^{i (n-1) \theta } \left(r f'(r)+n f(r)\right)}{r}+\\
&&\left( \frac{e^{i (n+1) \varphi } \left(r f'(r)-n f(r)\right)}{r}+F\right)\xi^\alpha.\label{trans1}\nonumber
\eea

Let us consider now the branches $F^{(2)}$ and $F^{(3)}$, since they can be considered a perturbation over the supersymmetric BbS model. Outside the defect, $r>R$ only survives the correction term in $\lambda$
\be
\delta\psi^\alpha\vert_{r\geq R}=\frac{\lambda}{2}\frac{\bar{W}'}{g(u,\bar{u})+2 h(u,\bar{u})u_{,\mu}\bar{u}^{,\mu}}\xi^\alpha.\label{transferex}
\ee

Now we take the symmetric potential (\ref{pot}). Once the solution satisfies the boundary conditions (\ref{bound}) the profile function must behave as follows close to the boundary
\be
f(r)=\beta \left(\frac{r-R}{R}\right)^\epsilon
\ee
for some $\epsilon>0$. The expansion of (\ref{transferex}) around the boundary of the defect leads to
\be
\delta\psi^\alpha\vert_{r\sim R}=\frac{\lambda  \bar{W}'\left(\bar{u}\right)}{4 \left(\epsilon ^2 \beta ^2 R^{-2 \epsilon } (r-R)^{2
   \epsilon -2}+\left(\beta  R^{-\epsilon } (r-R)^{\epsilon }\right)^s\right)}\xi^\alpha\label{tfer}
  \ee

We see from (\ref{tfer}) that the behavior of $\delta\psi^\alpha$ depends on the choice of the prepotential $W'$. Let us assume that $W'$ is a polynomial, and let $l$ be the minor degree of its monomials. Then
\be
W'(\bar{u})=m\bar{u}^l+\text{higher powers}
\ee
and $m$ a constant. 

\begin{table}
   \caption {SUSY transformation at the boundary} \label{tab} 
\begin{center}
\begin{tabular}{ l | c r }
   & $\lim_{r\rightarrow R}\delta\psi^\alpha$  \\ \hline
  $l>\min(2-\frac{2}{\epsilon},s)$ & $0$   \\ \hline
  $l<\min(2-\frac{2}{\epsilon},s)$& $\infty$ \\ \hline
    $l=\min(2-\frac{2}{\epsilon},s)$& $C\xi^\alpha$
\end{tabular}
\end{center}
\end{table}

The behavior of (\ref{tfer}) near the boundary of the compacton is summarized in Table I. In the first situation supersymmetry is broken inside the compacton but is restored outside (this situation corresponds to the one studied for the usual model if we consider non-BPS solutions), the consequence of this is the confinement of the fermionic zero modes to the defect. The second situation is forbidden since the transformation of the fermion diverges. The third situation is more interesting. Here the constant $C$ can take two values

\be
C= \begin{cases} 
    m\lambda\frac{ R^2   \beta ^{-2/\epsilon }}{4 \epsilon ^2}e^{-iln\varphi }, & \min(2-\frac{2}{\epsilon},s)=2-\frac{2}{\epsilon}\\
      m\frac{\lambda}{4}e^{-iln\varphi }&  \min(2-\frac{2}{\epsilon},s)=s.
   \end{cases}\label{case} 
\ee

Since the values given in (\ref{case}) are those taken at the boundary $r=R$, continuity of the SUSY transformation in $\mathbb{R}^2$ implies
\be
\delta\psi^\alpha\vert_{r\geq R}=C\neq 0.\label{fermdec}
\ee

The consequence of this fact is that SUSY is not only broken inside the defect but in all space, and therefore, the fermionic zero modes are no longer confined to the defect.

%%%%%%%%%%%%%%%%%%%%%%%%%%%%%%%%%%%%%%%%%%%
%%%%%%%%%%%%%%%%%%%%%%%%%%%%%%%%%%%%%%%%%%%
%%%%%%%%%%%%%%%%%%%%%%%%%%%%%%%%%%%%%%%%%%%
%%%%%%%%%%%%%%%%%%%%%%%%%%%%%%%%%%%%%%%%%%%
%%%%%%%%%%%%%%%%%%%%%%%%%%%%%%%%%%%%%%%%%%%
%%%%%%%%%%%%%%%%%%%%%%%%%%%%%%%%%%%%%%%%%%%
%%%%%%%%%%%%%%%%%%%%%%%%%%%%%%%%%%%%%%%%%%%
%%%%%%%%%%%%%%%%%%%%%%%%%%%%%%%%%%%%%%%%%%%
%%%%%%%%%%%%%%%%%%%%%%%%%%%%%%%%%%%%%%%%%%%

\section{Summary}

In this work we have studied the supersymmetric version of the BbS model. After a brief introduction to the superfield formulation of the model, we have discussed the relation between BPS equation and preservation of supersymmetry. We have found that generic BPS solutions preserve $1/4$ of supersymmetry, and therefore only one out of four supersymmetric generators remains unbroken. 

The fermionic sector of the model is rather cumbersome and contains derivatives of the auxiliary field. This fact makes it difficult to solve the fermionic equations of motion exactly. In order to circumvent this difficulty we have used  the supersymmetric transformations to obtain the fermionic zero modes. For BPS baby Skyrmions with compact support we have shown that fermionic zero modes only exist inside the defect, while outside $N=2$ supersymmetry is restored. However, if we add a D-term, we have shown that, even for compact (not necessarily BPS) solutions, all supersymmetric generators are broken in $\mathbb{R}^2$ for an appropriate choice of the prepotential and therefore, fermionic zero modes are not confined to the defect. This result is analogous to that obtained in \cite{Trodden} in four dimensions for supersymmetric cosmic strings with higher derivative terms. 

We have also presented a general framework to obtain $N=2$ versions from bosonic theories based on a method first described in \cite{Nitta1} for $N=1$ four dimensional theories and applied this result to the full bS model (Appendix B).
 
The generalization of these results to supersymmetric Skyrme like models in four dimensions and the study of the corresponding index theorems relating zero modes is under current investigation.

{\bf Acknowledgements.}- The author would like to thank Profs. C. Adam
and A. Wereszczynski for useful comments and Dr. V. Emelyanov for reading the manuscript.

%%%%%%%%%%%%%%%%%%%%%%%%%%%%%%%%%%%%%%%%%%%
%%%%%%%%%%%%%%%%%%%%%%%%%%%%%%%%%%%%%%%%%%%
%%%%%%%%%%%%%%%%%%%%%%%%%%%%%%%%%%%%%%%%%%%
%%%%%%%%%%%%%%%%%%%%%%%%%%%%%%%%%%%%%%%%%%%
%%%%%%%%%%%%%%%%%%%%%%%%%%%%%%%%%%%%%%%%%%%
%%%%%%%%%%%%%%%%%%%%%%%%%%%%%%%%%%%%%%%%%%%
%%%%%%%%%%%%%%%%%%%%%%%%%%%%%%%%%%%%%%%%%%%
%%%%%%%%%%%%%%%%%%%%%%%%%%%%%%%%%%%%%%%%%%%
%%%%%%%%%%%%%%%%%%%%%%%%%%%%%%%%%%%%%%%%%%%
\appendix

\section{Notation}

In this appendix we introduce our notation and conventions for $N=2$ supersymmetry in three dimensions. We work with the mostly minus metric. We have four  Grassmann variables, two chiral, $\theta^\alpha$ and two antichiral $\bar{\theta}^{\dot{\alpha}}$, from which it is possible to construct the superderivatives:
\be
D_\alpha=\frac{\pa}{\pa\theta^\alpha}+i\sigma^\mu_{\alpha\dot{\alpha}}\bar{\theta}^{\dot{\alpha}}\pa_\mu\quad \text{and}\quad \bar{D}_{\dot{\alpha}}=-\frac{\pa}{\pa\bar{\theta}^{\dot{\alpha}}}-i\theta^\alpha \sigma^\mu_{\alpha\dot{\alpha}}\pa_\mu.
\ee

The supersymmetric generators are equal to the superderivatives up to a sign
\be
Q_\alpha=\frac{\pa}{\pa\theta^\alpha}-i\sigma^\mu_{\alpha\dot{\alpha}}\bar{\theta}^{\dot{\alpha}}\pa_\mu\quad \text{and}\quad \bar{Q}_{\dot{\alpha}}=\frac{\pa}{\pa\bar{\theta}^{\dot{\alpha}}}-i\theta^\alpha \sigma^\mu_{\alpha\dot{\alpha}}\pa_\mu.
\ee

The only nonvanishing anticommutators among supercharges and superderivatives are the following
\be
\{D_\alpha,\bar{D}_{\dot{\alpha}} \}=-2i\sigma^\mu_{\alpha\dot{\alpha}}\pa_\mu\quad \text{and}\quad \{Q_\alpha,\bar{Q}_{\dot{\alpha}} \}=-2i\sigma^\mu_{\alpha\dot{\alpha}}\pa_\mu.
\ee

We chose the following Pauli matrices:
\be
\sigma^0=\left(\begin{matrix}
    1 &0   \\
     0&1
\end{matrix}\right), \quad
\sigma^1=\left(\begin{matrix}
    0 &1   \\
     1&0
\end{matrix}\right)\quad \text{and}\quad 
\sigma^2=\left(\begin{matrix}
    1 &0   \\
     0&-1
\end{matrix}\right).
\ee

In three dimensions and with two supersymmetries it is possible to construct chiral and antichiral superfields. A chiral superfield satisfies the constraint $\bar{D}_{\dot{\alpha}} \Phi=0$  and an antichiral one $D_\alpha\bar{\Phi}=0$.  The expansion in components of superfields satisfying the previous constraints leads to 
\bea
\Phi&=&u+i\theta \sigma^\mu\bar{\theta}\pa_\mu u +\frac{1}{4}\theta\theta\bar{\theta}\bar{\theta}\square u+\sqrt{2}\theta \psi-\frac{i}{\sqrt{2}}\theta\theta\pa_\mu\psi\sigma^\mu\bar{\theta}+\theta\theta F\\
\bar{\Phi}&=&\bar{u}-i\theta\sigma^\mu\bar{\theta}\pa_\mu\bar{u}+\frac{1}{4}\theta\theta\bar{\theta}\bar{\theta}\square \bar{u}+\sqrt{2}\bar{\theta}\bar{\psi}+\frac{i}{\sqrt{2}}\bar{\theta}\bar{\theta}\theta\sigma^\mu\bar{\psi}+\bar{\theta}\bar{\theta}\bar{F}
\eea
where $u$ and $F$ are complex fields and $\psi$ a two component complex spinor (respect. $\bar{u}$, $\bar{F}$ and $\bar{\psi}$).

%%%%%%%%%%%%%%%%%%%%%%%%%%%%%%%%%%%%%%%%%%%
%%%%%%%%%%%%%%%%%%%%%%%%%%%%%%%%%%%%%%%%%%%
%%%%%%%%%%%%%%%%%%%%%%%%%%%%%%%%%%%%%%%%%%%
%%%%%%%%%%%%%%%%%%%%%%%%%%%%%%%%%%%%%%%%%%%
%%%%%%%%%%%%%%%%%%%%%%%%%%%%%%%%%%%%%%%%%%%
%%%%%%%%%%%%%%%%%%%%%%%%%%%%%%%%%%%%%%%%%%%
%%%%%%%%%%%%%%%%%%%%%%%%%%%%%%%%%%%%%%%%%%%
%%%%%%%%%%%%%%%%%%%%%%%%%%%%%%%%%%%%%%%%%%%
%%%%%%%%%%%%%%%%%%%%%%%%%%%%%%%%%%%%%%%%%%%

\section{Other supersymmetric versions}

It was shown in \cite{Nitta1} that any bosonic model with one supersymmetry in four dimensions consisting of a complex scalar field  possesses a supersymmetric extension. It is immediate to extend this result to $N=2$ supersymmetry in three dimensions. The idea is quite simple and it is based on the tricky structure of the quartic term in superderivatives. Let us start with the following $N=2$ supersymmetric Lagrangian
\be
\mathcal{L}=\int d^2\theta d^2\bar{\theta} K(\Phi,\Phi^\dagger)+\int d^2\theta d^2\bar{\theta}\,\Lambda\, D^\alpha\Phi D_\alpha \Phi \bar{D}^{\dot{\beta}}\Phi^\dagger \bar{D}_{\dot{\beta}}\Phi^\dagger\label{B1}
\ee
where $\Lambda$ is a function depending on the superfields and its space-time derivatives and $K$ is the K\"ahler potential. Let us define $\Lambda_0=\Lambda\vert_{\theta=\bar{\theta}=0}$. After the integration we can solve the equation of motion for F in terms of the unknown function $\lambda$. We obtain two solutions
\be
F=0\quad\text{and\quad}F\bar{F}=-\frac{g}{2\Lambda_0}-\pa_\mu u\pa^\mu\bar{u}.
\ee

Now we take one of the solutions, for example $F=0$ (the procedure is exactly the same if we take the other) and substitute it into the Lagrangian, now
\be
\mathcal{L}\vert_{\text{on-shell}}= g\,\pa_\mu u\pa^\mu \bar{u}+\Lambda_0 (\pa_\mu u)^2(\pa_\nu \bar{u})^2.
\ee

We can determine the function $\Lambda_0$ by forcing the on-shell Lagrangian  to be equal to whatever we want, for example, the Lagrangian corresponding to the bS model (with quadratic term) and potential $V(u,\bar{u})$. We obtain
\be
\Lambda_0= -\frac{V(u,\bar{u})+h(u,\bar{u})\left(\pa_\mu u\pa^\mu\bar{u}\right)^2-h(u,\bar{u})(\pa_\mu u)^2(\pa_\nu \bar{u})^2}{(\pa_\mu u)^2(\pa_\nu \bar{u})^2}.
\ee

where $h(u,\bar{u})=\frac{1}{\left(1+u\bar{u} \right)^4}$. Now from the form of (\ref{B1}) it is obvious that only the $\theta=\bar{\theta}=0$ component of $\Lambda$ survives in the bosonic sector, therefore, the function $\Lambda$ we are looking for is simply
\be
\Lambda= -\frac{V(\Phi,\Phi^\dagger)+h(\Phi,\Phi^\dagger)\left(\pa_\mu \Phi\pa^\mu\Phi^\dagger\right)^2-h(\Phi,\Phi^\dagger)(\pa_\mu \Phi)^2(\pa_\nu \Phi^\dagger)^2}{(\pa_\mu \Phi)^2(\pa_\nu \Phi^\dagger)^2}.
\ee

In consequence, with this choice of $\Lambda$ the Lagrangian (\ref{B1}) constitutes an $N=2$ version of the full baby Skyrme model in the branch $F=0$. We can repeat the same procedure in the nontrivial branch to obtain another nonequivalent $N=2$ version of the model.

%%%%%%%%%%%%%%%%%%%%%%%%%%%%%%%%
%%%%%%%%%%%%%%%%%%%%%%%%%%%%%%%%
%%%%%%%%%%%%%%%%%%%%%%%%%%%%%%%%
%%%%%%%%%%%%%%%%%%%%%%%%%%%%%%%%
%%%%%%%%%%%%%%%%%%%%%%%%%%%%%%%%
%%%%%%%%%%%%%%%%%%%%%%%%%%%%%%%%

\section{Compactons in the D-deformed model}

In this appendix we demonstrate the existence of solutions satisfying (\ref{fermdec}), i.e. compact solutions allowing the fermionic zero modes to be nonzero outside their support. The addition of the D-term to the baby Skyrme action preserves $N=2$ supersymmetry and therefore, we expect that the general form of the BPS equation (\ref{bpssusy}) still holds. Let us expand the solution (\ref{Daux2}) close to the vacuum ($u=0$) in the symmetric ansatz
\be
F^{(2)}=f'(r)+\frac{f(r)^s}{2f'(r)}+\frac{m \lambda}{2}\frac{f(r)^l}{f'(r)^2}+\text{(higher order terms)}\label{expcom}.
\ee
 The holomorphic derivative of the field $u$ takes the form
\be
\pa u=\frac{-n f(r)+r f'(r)}{r}\label{holo}.
\ee

We do not take care about phase factors in the expressions above since they can be absorbed in the arbitrary phase of  (\ref{bpssusy}) (and they do not change the r-dependence of the solutions). The  behavior of the solution close to the vacuum determines if the solution is compacton-like \cite{Arodz,Adamcom}. If such a solution approaches to the vacuum in a powerlike (positive) manner it has compact support. On the other hand, if it approaches to the vacuum exponentially or negative power-like, we have solutions with infinite support (exponentially localized or powerlike localized respectively). For simplicity we do not consider here the situation when $l=s$ (no compact solutions in this case). When $l\neq s$ we can split (\ref{expcom}) into two terms. Accordingly we can rewrite the BPS equation in the vicinity of the minimum of the potential as follows
\bea
-n\frac{f(r)}{r}&=&\frac{f(r)^s}{2f'(r)},\quad s<l\\
-n\frac{f(r)}{r}&=&\frac{m\lambda}{2}\frac{f(r)^l}{f'(r)^2},\quad s>l.
\eea

The solution close to the vacuum has the following r-dependence
\be
 f(r)\propto\begin{cases} 
      r^{\frac{2}{2-s}} & s< l \\
      r^{\frac{3}{3-l}} & s>l
   \end{cases}
\ee

In the first case we have compactons  for $s\in (0,2)$ while in the second for $l\in[0,3)$ (the $s=0$ is removed since implies a constant potential). Note that if $s<l$ the condition  $l=\min(2-\frac{2}{\epsilon},s)$ (Table I) implies that $l=s$ and therefore cannot be fulfilled. The consequence is that in this case, fermionic zero modes are confined to the support of the compactons. Now for $s>l$, the previous condition can be fulfilled for $l=0$, and therefore from (\ref{case}),
\be
\delta\psi^\alpha\vert_{r\geq R}=m\lambda \frac{R^2}{4\beta^2}\neq0.
\ee

The conclusion in that only linear prepotentials can generate compact solutions, such that the corresponding fermionic zero modes are not confined to the support of the compacton.

\end{document}